%% file: main.tex
\documentclass[conference]{IEEEtran}
\input{imports}

\input{macros}
\begin{document}

\title{\system: A Simulation Framework for Evaluating the Concrete Scalability of Secure Aggregation Protocols}

\author{\IEEEauthorblockN{Ivoline C. Ngong}
\IEEEauthorblockA{University of Vermont\\
kngongiv@uvm.edu}
\and
\IEEEauthorblockN{Nicholas Gibson}
\IEEEauthorblockA{University of Vermont\\
Nicholas.Gibson@uvm.edu}
\and
\IEEEauthorblockN{Joseph P. Near}
\IEEEauthorblockA{University of Vermont\\
jnear@uvm.edu}}

\maketitle

\begin{abstract}
Recent secure aggregation protocols enable privacy-preserving federated learning for high-dimensional models among thousands or even millions of participants. Due to the scale of these use cases, however, end-to-end empirical evaluation of these protocols is impossible. We present \system, a framework for empirical evaluation of secure protocols via simulation. \system provides an embedded domain-specific language for defining protocols, and a simulation framework for evaluating their performance. We implement several recent secure aggregation protocols using \system, and perform the first empirical comparison of their end-to-end running times. We release \system as open source.\end{abstract}

\section{Introduction}

\emph{Federated learning}~\cite{kairouz2021advances} allows for collaborative distributed training of machine learning models without requiring training data to be collected centrally. By keeping training data decentralized, federated learning can reduce privacy risks for individuals who contribute training data. However, recent work has shown that in some cases, the individual model updates computed during federated learning can reveal a surprising amount about the original training data.

To address this challenge, \emph{secure aggregation} protocols can be used to construct federated learning systems that reveal only \emph{aggregated} model updates, providing much stronger protection against privacy attacks. Combined with differential privacy~\cite{dwork2014algorithmic}, secure aggregation protocols can enable truly privacy-preserving federated learning. Recent work in secure aggregation has produced protocols that scale to high-dimensional model updates~\cite{bonawitz2017practical} and millions of clients~\cite{bell2020secure}; in theory, these approaches scale well enough to meet the requirements of industry-scale machine learning.

However, evaluating these protocols empirically remains a major challenge, because of the sheer scale of the use cases they are designed for. For example, evaluating a secure aggregation protocol with 10,000 clients is impossible for most researchers, because it requires provisioning 10,000 physical machines to perform the experiment. As a result, previous work has focused on evaluating individual components of a protocol in isolation~\cite{bonawitz2017practical} or simply reporting properties of the protocol's \emph{expected} performance by analyzing the protocol itself~\cite{bell2020secure}. Such evaluations are effective for comparing the asymptotic complexities of protocols, but may not capture the protocol's concrete performance.

We present \system, a simulation framework for the empirical evaluation of secure aggregation protocols. \system is designed to evaluate the concrete, \emph{end-to-end} performance of protocols \emph{at scale}, by leveraging an accurate simulation of hundreds or thousands of parties on a single machine. For example, \system can perform a simulation of the Bell et al.~\cite{bell2020secure} protocol for 10,000 clients on a single machine in just a few hours.

The \system framework provides a simulator that accurately measures the end-to-end running time of protocols, including both communication and computation time. To model computation cost in many-party protocols, \system records the actual computation time for each party, and simulates these computations running in parallel. To model communication cost, \system uses a model of network latency based on actual internet latency data collected from internet speed tests. \system builds on the existing \abides framework~\cite{byrd2019abides} to coordinate the simultaneous execution and communication of the parties and measure the total running time of the protocol.

To ease the implementation of new protocols, \system provides a domain-specific language (DSL) embedded in Python for defining synchronous secure aggregation protocols. The \system DSL makes it straightforward to translate protocol descriptions into implementations, and also provides utilities for common cryptographic constructs like public-key encryption and secret sharing. We have used the \system DSL to implement existing several protocols from the literature in fewer than 100 lines of code.

We use \system to conduct an empirical comparison between several existing protocols implemented in our case studies. The results are mostly consistent with existing conclusions about protocol performance, but also yield new insights about the concrete performance of these protocols. For example, we show that network latency has relatively little effect on the total running time for these protocols, and that packed secret sharing has a significant impact on performance for some protocols. \newtext{In addition, we empirically validate the simulator's accuracy by comparing it against ``ground truth'' performance results obtained by executing the same protocols on real hardware.}

We release the \system framework and our case study implementations as open source.\footnote{\url{https://github.com/uvm-plaid/olympia}} In addition to comparing the performance of existing protocols, we hope that \system will be useful as a standardized benchmarking tool for new protocols, and also as a tool for helping to refine existing protocols for better performance, to develop new protocols, and to evaluate protocol suitability for specific real-world deployment scenarios.

\paragraph{Contributions.}
In summary, our contributions are:
\begin{itemize}[leftmargin=10pt, itemsep=0pt,topsep=0pt]
\item We present \system, a simulation framework for secure aggregation protocols that accurately models the concrete performance of protocols with millions of participants
\item We use \system to evaluate several existing protocols at scales that are not practical without a simulator
\item We show that evaluation with \system leads to important insights about the concrete performance of protocols, including some that can help improve protocol performance
\item \newtext{We validate \system's simulation accuracy by comparing its results against actual execution times for a small number of clients}
\end{itemize}

\section{Overview of \system}

Traditional \emph{secure multiparty computation} (MPC)~\cite{evans2017pragmatic} protocols are designed to work best for a handful of parties---2- and 3-party computation are most common, and most protocols are evaluated with a single-digit number of parties. Some more recent protocols have been evaluated using as many as 128 parties~\cite{wang2017global}. At this scale, empirical evaluation is possible: a separate physical machine can be used for each party, allowing realistic measurement of total running time.

\emph{Secure aggregation protocols} are designed to work for much larger sets of parties---typically, hundreds to thousands (or even millions). At this scale, experimental evaluation is not practical; it's simply not feasible to provision enough machines. Instead, authors typically report computation and communication \emph{complexity} as a proxy for experimental results~\cite{bell2020secure}; in some cases, authors additionally implement the protocol and measure concrete computation time for a \emph{single} client~\cite{bonawitz2017practical} and for a server with fixed client inputs.

For secure aggregation protocols, authors typically \textbf{do not report end-to-end wall-clock time} resulting from an experimental evaluation, because it is not feasible to run such an experiment~\cite{bonawitz2017practical,bell2020secure,so2021turbo, kadhe2020fastsecagg, stevens2022efficient, yang2021lightsecagg, guo2022microfedml, roth2019honeycrisp,roth2020orchard}. As we discuss later, the inability to measure concrete performance of these protocols makes it difficult to understand their relative performance properties.

\paragraph{\system: Evaluation via Simulation.}
The goal of this work is to enable experimental evaluation of secure aggregation protocols \emph{at scale}. Experimental evaluation with \system can highlight surprising mismatches between analytic bounds and concrete performance and can suggest simple methods for significantly improving protocol performance. In addition, \system can simplify deployment, by enabling developers to predict performance in advance.

\paragraph{Challenge: realism \& scale.}
The primary challenge lies in building a framework that works at scale and produces accurate results. \system is built on \abides~\cite{byrd2019abides}---a simulation framework originally designed for high-frequency trading applications in financial markets. \abides is designed to simulate a large number of \emph{agents} running simultaneously and communicating asynchronously. \abides simulates concurrent execution of these agents with high precision and accuracy.
\abides has previously been used to evaluate the performance of specific secure aggregation protocols~\cite{byrd2022collusion, guo2022microfedml, byrd2020differentially}. 
To support the evaluation of secure aggregation protocols, \system adds cost models for network traffic (based on real-world latency data and configurable bandwidth limitations) and computation (based on actual execution time), and a DSL that simplifies the specification of new protocols.


\paragraph{New Insights from \system.}
By enabling empirical analysis at scale, \system can lead to important insights about concrete protocol performance. We implemented several state-of-the-art secure aggregation protocols in \system (Section~\ref{sec:case-studies}); our experimental evaluation of these protocols (Section~\ref{sec:evaluation}) confirms their expected asymptotic behavior, but also surfaces performance properties not obvious from the asymptotics alone. For example, the results suggest that computation time has a much larger effect on total running time than network latency or round complexity; in addition, for one protocol, our results demonstrate the practical importance of optimizations like packed secret sharing that improve concrete performance without changing computation or communication complexity.



\section{The \system Framework}

\system provides two main components: a domain-specific language (DSL) for describing single-server secure aggregation protocols, and a simulation framework for evaluating the practical concrete performance of these protocols.

\paragraph{The \system DSL.}
As a DSL, this framework could potentially be used for extracting a protocol's implementation, evaluating its performance based on round complexity, computational complexity, bandwidth cost, and hyper parameter impact, etc. We describe the \system DSL in Section~\ref{sec:system-dsl}.

\paragraph{The \system Simulator.}
Its implementation is based on the discrete event simulation framework \abides, but implements the round-based synchronous communication commonly used in secure aggregation protocols. In addition, the \system simulator is specifically designed to accurately record both computation and communication cost, by using realistic models for both components. We describe the \system simulator in Section~\ref{sec:system-simulator}.

\subsection{The \system DSL}
\label{sec:system-dsl}
Protocols in \system are defined in three parts: a \emph{server class} implementing the aggregation server's behavior, a \emph{client class} implementing the client's behavior, and a \emph{config file} that determines protocol setup and parameters. Multiple experiments can be quickly reconfigured and run with varying simulation parameters using a single setup. 

The server and client classes implement what each party does in each round. Clients and servers are defined in \system by inheriting from the \texttt{AggregationClient} and \texttt{AggregationServer} classes, which implement the minimal set of properties and methods necessary for efficient communication between agents and interaction with simulation kernels. Client and server classes override methods to define the behavior of the protocol. Table~\ref{tbl:API} describes the \system API in terms of the key methods used in defining a protocol.
\begin{table}
\centering
\renewcommand{\arraystretch}{1.5}
\begin{tabular}{|p{0.13\textwidth}|p{0.23\textwidth}|}
\hline 
\textbf{Method} & \textbf{Description} \\
\hline
\texttt{round()}: $\textit{round \#} \times \textit{message(s)} \rightarrow \textit{next message(s)}$ \textbf{(server and client)} & The server is given the round number and incoming messages from the agents and implements what takes place in each round. \\
\hline 
\texttt{next\_round()}: $\textit{round \#} \times \textit{message(s)} \rightarrow \textit{bool}$ \textbf{(server only)} & Given the current round number and incoming messages from the clients, determines whether it is time to move to the next round. \\
\hline 
\texttt{succeed()} \phantom{hi} \textbf{(server only)} & Called at the end of the protocol, to indicate success and record results \\
\hline 
\texttt{fail()} \phantom{hi hi} \textbf{(server only)} & Called during the protocol to indicate failure \\
\hline 
\end{tabular}
\caption{\system API (server and clients).}
\label{tbl:API}
\end{table}

\paragraph{Example: Baseline Insecure Summation Protocol.}
Here, we present a simple example that allows clients to submit their inputs to a \emph{trusted} server, which computes the sum of these inputs. Though this protocol is \emph{insecure}, it serves as a good example for how protocols are defined in \system, and a useful baseline for performance comparisons in our evaluation.


After setting up all required parameters in the baseline configuration file, the protocol is implemented as shown in Figure~\ref{fig:baseline_impl_client}. The server for this protocol (and for the other protocols we implement later) inherit from the \inline{DropoutAggregationServer}, which simulates a fraction of clients dropping out during each round of the protocol.

\begin{figure} 
\begin{minted}[linenos,xleftmargin=0pt,
numbersep=4pt,fontsize=\scriptsize]{python}
class BaselineClient(AggregationClient):
  def round(self, round_number, message):
    if round_number == 1:   # send secret input to server
      return self.GF(self.secret_input)

class BaselineServer(DropoutAggregationServer):
  def round(self, round_number, messages):
    if round_number == 1:   # start the protocol
      self.threshold = int(len(self.clients) * .95)
      return {client: None for client in self.clients}

    elif round_number == 2: # sum up received vectors
      self.succeed(GF(list(messages.values())).sum(axis=0))
\end{minted}
\caption{\system implementation of the baseline protocol.}
\label{fig:baseline_impl_client}
\end{figure}

\subsection{The \system Simulator}
\label{sec:system-simulator}

The \system simulator is based on \abides, a Python application built around actors called agents, who communicate via \emph{asynchronous} messages. In the \abides framework, agents are instances of subclasses that inherit from the base \texttt{Agent} class. Subclasses implement what each agent does in a protocol, with the same type of agents having only one subclass. For instance, it is sufficient to create two subclasses for a protocol with a single server and multiple clients, although each client can have different attributes like timing. 
The base Agent class in \abides implements the minimum required methods for each agent to properly interact with the simulation kernel.


\abides lacks important functionalities for our setting, which \system's simulator implements: measuring communication costs and computation time, modeling a WAN network, simulating dropouts realistically, and evaluating how other significant parameters (input vector size, finite field size, etc.) affect the protocol.

\paragraph{Overview of simulation approach.}
\newtext{The \system simulator runs the protocol by executing its components sequentially, passing messages between the parties as the protocol specifies. The simulator measures the execution time of each component individually, and calculates total running time by \emph{simulating} the parallelism of actual protocol execution. When two component executions could run in parallel (because neither depends on a message from the other---e.g. when two clients compute their responses to a message from the server), then the simulator calculates the total time of both components as the maximum of their execution times (rather than the sum). This approach allows simulating protocols with thousands of parties on a single computer, while accurately modeling the parallelism that would occur in actual execution of the protocol.}

\paragraph{Modeling computation time.}
\system models computation delays by measuring the actual time the protocol takes to complete the computation, and accounting for that time as computation time spent by the appropriate party in the protocol. 
In \abides, computation time is measured by asking the programmer to specific explicit \emph{computation delays}, and tracking each party’s individual time using these delays as the simulation progresses. 

In \system, on the other hand, the simulator measures actual computation time for each party. The simulation runs each party's computation sequentially, and then aggregates the computation delays to simulate the parallel execution of client computations. While this enables us to capture realistic computation times, it also means that accurate results depend on the actual efficiency of the protocol's implementation. In our case study implementations, we use efficient libraries to implement computationally challenging features (e.g. PyNaCl for public-key encryption; Galois for finite field operations).



\paragraph{Modeling network latency.}
\abides supports the modeling of communication latency between different parties in the protocol, which is crucial to creating realistic simulations. In this model, a two-dimensional \emph{latency matrix} defines the minimum nanosecond delay between each pair of parties. The kernel uses this matrix and a noise model to simulate network conditions. 

\system models network delays using a real-world internet speed test dataset~\cite{speed_test} to simulate deployment on a wide-area network. Based on zoom level 16 web mercator tiles (approximately 610.8 meters by 610.8 meters at the equator), the dataset provides global fixed broadband and mobile (cellular) network latency measurements. 
The latency matrix is computed by measuring the latency between each endpoint and their respective internet providers, then adding the speed-of-light latency between their geographic locations.

\paragraph{Modeling network bandwidth limitations.}
In addition to modeling network latency, \system also provides added support for modeling network bandwidth limitations. Bandwidth limits can be configured for both the clients and the server, and are incorporated into the total running time measured by the simulator. The default setting in \system is no bandwidth limit; adjusting the limit can be especially useful to enable modeling of situations where the server has limited bandwidth, as described in our evaluation.

\paragraph{Modeling dropouts.}
Many secure aggregation protocols (including the case studies described in Section~\ref{sec:case-studies}) are robust to a fraction of the clients dropping out during the execution of the protocol. In some cases, clients dropping out can affect protocol performance, so \system is capable of simulating dropouts in order to surface these effects. \inline{DropoutAggregationServer} represents a server intended to be robust against dropouts; implementations specify an upper bound $\delta$ on the expected fraction of dropouts, and the server moves on to the next round after it has received at least $n\delta$ messages from clients.

\paragraph{Checking correctness.}
\system is not designed to verify the correctness of protocol implementations, but it can be used as a tool for end-to-end testing of protocols. Our case studies are implemented to facilitate this kind of testing, by providing consistent inputs on each run of the protocol and checking that the protocol's output is as expected. \system's simulator can be used to test correctness under varying conditions that may surface bugs---for example, we found and fixed several bugs in our implementations by varying the number of clients and the simulated dropout rate.

\paragraph{Threat model.}
Our case study protocols can be implemented with either semi-honest or malicious security. \system does not actually simulate malicious parties, and is not designed to evaluate security. However, \system can be used to evaluate the additional overhead of malicious-secure protocol variants. For our case studies, we implemented both semi-honest and malicious-secure variants, and our evaluation includes a comparison of the associated overhead.

\paragraph{Libraries and cryptographic primitives.}
\newtext{\system is primarily Python-based, and integrates most easily with other Python libraries. To produce high-performance, practically secure protocol implementations, most protocol implementations will leverage Python wrappers around efficient, well-tested libraries (usually implemented in C). For example, our case studies use PyNaCl (a wrapper around libsodium) and Galois (a finite field library that uses BLAS/LAPACK for array operations). Usage of well-tested libraries for cryptographic building blocks can also prevent subtle side-channel vulnerabilities.}


\paragraph{Configuration and experiments.}
Experiments in \system are configured using a YAML file that specifies the parameters to vary (e.g. the number of clients, the dimensionality of the inputs, and other protocol parameters). Given a configuration file, \system runs the specified experiments on a single machine and saves the results to a CSV file. \system is capable of running experiments involving tens of thousands of clients and vectors with millions of elements on a single machine (our experiments required only 64GB of memory).


\subsection{Supported Protocol Designs}

\system is designed to support a variety of protocol types and architectures---the only restriction is that the protocol must be synchronous and proceed in rounds. Protocol implementations can contain arbitrary Python code, and messages between parties can contain any Python value whose size in bytes can be measured. Though our case studies focus on single-server secure aggregation protocols, \system's flexibility makes it suitable for evaluating many other protocols as well.

\paragraph{Single-server secure aggregation.}
Our case studies focus on single-server secure aggregation protocols, because protocols in this category are often designed for hundreds or thousands of parties and are therefore especially difficult to evaluate by other means. \system provides API support for this setting, by defining \inline{AggregationServer} and \inline{AggregationClient} classes to be extended by protocol implementations.

\paragraph{Protocols involving multiple servers.}
\system can also support multi-server protocols like Prio~\cite{corrigan2017prio} and its descendants. Implementing multi-server protocols requires defining a subset of the parties to be servers, and defining a class that specifies the servers' behavior, and then defining the clients to send messages directly to the multiple servers. \newtext{This setting requires clients to send messages to multiple servers; to implement it in \system, the programmer would need to implement new client and server classes with explicit \inline{send} operations to specify the communication pattern.}

\paragraph{Peer-to-peer protocols.}
\system also supports peer-to-peer protocols, including general MPC protocols that evaluate circuits. These protocols are not typically designed to scale to thousands of parties, and can often be evaluated empirically without \system, so our case studies do not focus on protocols of this type. Recent work designed to scale to larger numbers of parties~\cite{gordon2021more} may benefit from evaluation using \system. \newtext{This setting requires clients to send messages to each other; like the multi-server setting, the programmer would need to specify each \inline{send} operation manually to implement it.}

\paragraph{Input validation.}
Recent approaches for secure aggregation with input validation~\cite{corrigan2017prio, roy2022eiffel, bell2022acorn} can also be evaluated using \system. These approaches typically add a layer on top of an existing protocol, and do not change the protocol's communication patterns; as long as the additional layer can be implemented using Python, it can be evaluated in \system.

\paragraph{Threat models.}
\system does not evaluate security, and it is agnostic to the protocol's intended threat model. Variants of protocols that target different threat models (including all possible combinations of corrupted and honest parties) can be implemented and compared in \system. Our evaluation includes semi-honest and malicious-secure variants of several protocols.

\newtext{
\paragraph{Limitations.}
\system is designed mainly for single-server, synchronous protocols and although it can support any synchronous protocol, its use is limited in asynchronous settings. Since it is based on Python, integrating code from other programming languages can be challenging, limiting its flexibility. Additionally, the fact that computations are executed sequentially for each client also means that experiments can be time-consuming, particularly for more complex protocols. Moreover, implementing peer-to-peer protocols demands extra coding effort, as \system's DSL is primarily oriented towards single-server and client-server protocols.
}
\section{Case Studies}
\label{sec:case-studies}

This section describes our implementations of six secure aggregation
protocols in \system. The simplest of these uses secret sharing to add
the input vectors (\S~\ref{sec:secret-sharing}); it requires $O(nl)$
per-client communication for $n$ clients and vectors of length $l$, so
it is not practical for large $n$ or large $l$.
The protocols of Stevens et al.~\cite{stevens2022efficient}
(\S~\ref{sec:stevens}) and Bonawitz et
al.~\cite{bonawitz2017practical} (\S~\ref{sec:bonawitz}) improve
per-client communication cost to $O(n + l)$; these protocols work well
for large vectors, but not for huge sets of clients. Finally, the Bell
et al.~\cite{bell2020secure} (\S~\ref{sec:bell}), Sharing
sharing~\cite{stevens2022secret}, and ACORN~\cite{bell2022acorn}
protocols improve per-client communication cost even further---to $O(
\log n + l)$---making them suitable for large sets of clients. Table~\ref{tbl:casestudies} summarizes our case studies.

\begin{table*}
\centering
\renewcommand{\arraystretch}{1.5} 
\begin{tabular}{|l|l|l|l|l|l|l|}
  \hline
  \textbf{Protocol} & \textbf{Setting} & \multicolumn{2}{l|}{\textbf{Client}} & \multicolumn{2}{l|}{\textbf{Server}} & \textbf{\S}\\
  & & \textbf{Communication} & \textbf{Computation} & \textbf{Communication} & \textbf{Computation} & \\
  \hline
  Secret sharing & --- & $O(ln)$ & $O(ln)$ & $O(ln)$ & $O(ln log(n))$ & \ref{sec:secret-sharing}\\
  \hline
  Stevens et al.~\cite{stevens2022efficient} & Few clients  & $O(l+k+n)$ & $O(lk+nlogn)$ & $O(ln+k)$ & $O(lk+ln+nlogn)$ & \ref{sec:stevens}\\
  \hline
  Bonawitz et al.~\cite{bonawitz2017practical} & Few clients & $O(n+l)$ & $O(n^2+nl)$ & $O(n^2+nl)$ & $O(ln^2)$ & \ref{sec:bonawitz}\\
  \hline
  Bell et al.~\cite{bell2020secure} & Many clients & $O(logn+l)$ & $O(log^2n+llogn)$ & $O(nlogn+nl)$ & $O(nlog^2n+nllogn)$ & \ref{sec:bell}\\
  \hline
  Sharing sharing~\cite{stevens2022secret} & Many clients & $O(l+logn)$ & $O(l+log^2n)$ & $O(ln)$ & $O(ln)$ & \ref{sec:sharingsharing}\\
  \hline
  ACORN~\cite{bell2022acorn} & Many clients & $O(l+logn)$ & $O(l+logn)$ & $O(n(l+logn))$ & $O(n(l+logn))$ &\ref{sec:acorn}\\
  \hline
\end{tabular}
\caption{\newtext{Communication and computation complexities of case study protocols implemented in \system,  for n parties aggregating vectors of size l and LWE security parameter k}}
\label{tbl:casestudies}
\end{table*}


\subsection{Common Elements}
\label{sec:common-elements}

This section describes some common elements used in all of the
protocols implemented in our case studies.

\paragraph{Principle: peer-to-peer communication via the server.}
The \system API is designed for building single-server aggregation
protocols, in which clients communicate only with the server. This
setup is designed to model real-world constraints on federated
learning systems, in which clients are often mobile devices that may
be protected by a firewall, and clients may have limited connectivity.
However, many secure aggregation protocols require clients to send
messages to each other, and require hiding the contents of those
messages from the server. To accomplish this, single-server secure
aggregation protocols typically use public-key cryptography to allow
the clients to communicate \emph{through} the server, and the server
simply routes messages to the correct clients. All of the protocols
described in this section make use of this approach.

\paragraph{Building block: public-key encryption.}
We make the same assumptions about public-key cryptography as Bonawitz
et al.~\cite{bonawitz2017practical}. Specifically, the protocols
described in this section rely on Diffie-Hellman key
agreement~\cite{diffie1976new}, which describes a \emph{key
  generation} procedure to generate public and private keys, and a
\emph{key agreement} procedure that combines party $a$'s private key
with party $b$'s public key to form a shared symmetric encryption
key---and produces the \emph{same} symmetric key when performed in
reverse (i.e. $\textbf{agree}(sk_a, pk_b) = \textbf{agree}(sk_b,
pk_a)$).
In our implementations, we implement public-key cryptography using
PyNaCl, a wrapper around the efficient libsodium
library.

\paragraph{Building block: threshold secret sharing.}
A $(t, n)$ \emph{threshold secret sharing} scheme allows a party to
split a secret input into $n$ \emph{shares}, such that each individual
share reveals nothing about the secret input, but $t$ shares enable
reconstruction of the secret. Such schemes also have an \emph{additive
  homomorphism}---they allow adding shares of \emph{different} secrets
together to compute one share of the sum of the secrets. The most commonly-used
example is Shamir's secret sharing scheme~\cite{shamir1979share}.
%

\system provides an efficient implementation of Shamir secret sharing,
leveraging the Galois library for working with finite
fields.
We also provide an implementation of \emph{packed secret sharing},
which encodes more than one secret in a single share. This scheme adds
a parameter $k$ to the \texttt{share} function; it encodes $k$ field
elements in a single share, but requires at least $t + k$ shares for
reconstruction. Packed secret sharing can improve concrete performance
in some protocols, as our evaluation shows.

\subsection{Secret Sharing Protocol}
\label{sec:secret-sharing}

Our first case study is a simple protocol that uses threshold secret
sharing to perform secure aggregation. This protocol was designed as a
simple secure baseline, and only performs well when both the number of
clients ($n$) and the size of the aggregated vectors ($l$) are small.

\begin{figure}
  \begin{minted}[linenos,xleftmargin=0pt, numbersep=4pt,fontsize=\scriptsize]{python}
class SecretSharingClient(AggregationClient):
  def round(self, round_number, message):
    if round_number == 1:   # generate keys
      self.sk_u = PrivateKey.generate()
      return self.sk_u.public_key
    elif round_number == 2: # generate shares
      self.pks, n = message, len(message)
      shares = shamir.share_array(self.secret_input,
                                  n, n//2)
      return {c: shares[c].encrypt(self.sk_u, pk)
              for c, pk in self.pks.items()}
    elif round_number == 3: # sum up received shares
      dec_shares = [s.decrypt(self.sk_u, self.pks[c])
                    for c, s in message.items()]
      return shamir.sum_share_array(dec_shares, axis=0)
  \end{minted}
  \caption{\system implementation of the secret sharing protocol (client).}
  \label{fig:shamir_impl_client}
\end{figure}

\begin{figure}
  \begin{minted}[linenos,xleftmargin=0pt, numbersep=4pt, fontsize=\scriptsize]{python}
class SecretSharingServer(DropoutAggregationServer):
  def round(self, round_number, messages):
    if round_number == 1:   # start the protocol
      return {client: None for client in self.clients}
    elif round_number == 2: # broadcast public keys
      return {client: messages for client in self.clients}
    elif round_number == 3: # route shares to clients
      return route_messages(messages)
    elif round_number == 4: # reconstruct sum
      vs = messages.values()
      self.succeed(shamir.reconstruct_array(vs))
  \end{minted}
  \caption{\system implementation of the secret sharing protocol (server).}
  \label{fig:shamir_impl_server}
\end{figure}

\paragraph{Protocol description.}
For $n$ clients and one server, aggregating vectors of field elements
with length $l$, the high-level idea of the protocol is as follows:
first, each client sends one share of its input to each other client;
second, each client sums the shares it receives to compute one share
of the total sum, and sends this share to the server; third, the
server uses the shares to reconstruct the sum.
Since each client needs to generate $n$ shares for each of $l$ vector
elements, the per-client communication cost is $O(nl)$.

\paragraph{Protocol implementation.}
The complete \system implementation of this protocol appears in
Figure~\ref{fig:shamir_impl_client} (client) and
Figure~\ref{fig:shamir_impl_server} (server). The implementation
proceeds as follows:
\begin{itemize}[leftmargin=12pt, itemsep=0pt]
\item \textbf{Round 1}: Each client broadcasts their public key
  (client lines 4-5).
\item \textbf{Round 2}: Each client $P_i$ generates $n$ shares with
  threshold $t$ of each element of their input vector $x_i$. $P_i$
  sends one share to each other client $P_j$ (and keeps one for
  itself) (client lines 8-9).
\item \textbf{Round 3}: Each client $P_i$ receives 1 share of each
  other client $P_j$'s input. $P_i$ adds these shares together, to get
  one share of the sum of all clients' inputs. $P_i$ sends the share
  of the sum to the server (client lines 12-14).
\item \textbf{Round 4}: The server receives $n$ shares of the total
  sum of the inputs, and reconstructs the total sum (server line 10).
\end{itemize}



\subsection{Stevens et al. Protocol}
\label{sec:stevens}

Stevens et al.~\cite{stevens2022efficient} design a protocol with
similar structure to the secret sharing protocol, but leverage the
learning with errors (LWE) assumption~\cite{regev2009lattices} to reduce the
dimensionality of the vector being secret shared to (effectively) a
constant that depends on the security parameters.

The LWE assumption says that for a public matrix $A \in
\mathbb{F}_q^{l \times s}$, secret vector $S \in \mathbb{F}_q^s$, and
secret \emph{error} vector $E \in \chi^l$, it is computationally hard
to distinguish the pair $(A, B)$ from a pair of uniformly random
numbers, where  $B = A \cdot S + E$,
even when $s \ll l$.

\paragraph{Protocol description.}
The insight behind the Stevens et al. protocol is to use the vector $B
\in \mathbb{F}_q^{l}$ as a \emph{random mask} to send the secret input
vector to the server, but aggregate the shorter $S$ vector using
secret sharing instead. This reduction in dimensionality leads to a
corresponding reduction in communication cost.
%

\begin{figure}
  \begin{minted}[linenos,xleftmargin=0pt, numbersep=4pt,fontsize=\scriptsize]{python}
class StevensClient(SecretSharingClient):
  def round(self, round_number, message):
    if round_number == 2:
      S = self.GF.Random(self.s_len)
      e = gen_noise()
      A = GF.Random((self.dim, self.s_len), seed=seed)
      masked_value = self.secret_input + A.dot(S) + e
      self.secret_input = S  # secret share S vector
      return self.masked_value, 
             ShamirClientAgent.round(self, round_number, 
                                           message)
    else:
      return ShamirClientAgent.round(self, round_number, 
                                     message)
  \end{minted}
  \caption{\system implementation of the Stevens et
    al.~\cite{stevens2022efficient} protocol (client).}
  \label{fig:stevens_impl_client}
\end{figure}

\paragraph{Protocol implementation.}
The structure of the protocol matches that of the secret sharing
protocol (Section~\ref{sec:secret-sharing}) exactly. Because of the
similarities, we can leverage the secret sharing protocol in our
implementation, by subclassing the secret sharing protocol and
modifying only round 2 (for the client) and round 4 (for the server).
The client implementation appears in
Figure~\ref{fig:stevens_impl_client}. In round 2, the clients generate
a random $S_i$ (line 4) and secret share it (instead of their input
vector $x_i$; line 7), and also send their masked vector $x_i + B_i$
to the server (line 8). In round 4, the server subtracts the
aggregated masks after reconstructing $\sum_i S_i$. No other changes
are necessary.


\subsection{Bonawitz et al. Protocol}
\label{sec:bonawitz}

Bonawitz et al.~\cite{bonawitz2017practical} design a protocol with
per-client communication cost of $O(n + l)$. The key idea behind this
approach is to leverage \emph{pairwise masking}: if client $A$ adds a
random \emph{mask} to their input vector, and client $B$ subtracts the
\emph{same} mask from their input vector, then summing up the two
masked vectors eliminates the mask and yields the sum of the original
vectors.

In the Bonawitz et al. protocol, each client adds such a mask to their
vector for \emph{each other client}. Adding up all of the masked
vectors yields the sum of the original input vectors, and the server
is prevented (by the masks) from learning any of the individual input
vectors. To achieve reduced communication cost, clients agree on their
pairwise masks by computing them from each other's public keys, by
using the result of the $\textit{agree}$ function
(Section~\ref{sec:common-elements}) as the random seed for a
pseudorandom number generator (PRNG) and using the PRNG to generate
the mask vector.

%
We implemented the semi-honest and malicious variants of the Bonawitz
et al. protocol in \system. Our implementation uses the same Shamir
secret sharing primitives and public-key encryption as the secret
sharing baseline (Section~\ref{sec:secret-sharing}). Per-client
communication cost for the protocol includes the masked vector itself
($O(l)$), plus the client's public key ($O(1)$) and the Shamir shares
of $b_u$ and $s_{u,v}$ ($O(n)$).

\subsection{Bell et al. Protocol}
\label{sec:bell}

The communication complexity of the Bonawitz et al. protocol is nearly
optimal for small sets of clients and large vectors, but its
per-client communication cost is much higher than an insecure solution
when $n$ is large. The Bell et al. protocol~\cite{bell2020secure}
modifies the Bonawitz et al. protocol by relaxing the requirement that
each client add a mask for {each other client}. Instead, in the Bell
et al. protocol, each client adds a mask for a \emph{subset} of the
other clients.

The Bell et al. protocol generates a graph connecting clients to a set
of $k$ \emph{neighbors} with whom they will exchanged pairwise masks.
By setting $k = \log n$, the protocol reduces the number of masks in
each masked vector from $O(n)$ to $O(\log n)$, and thus also reduces
the per-client communication cost.
The original Bonawitz et al. protocol can be recovered from the Bell
et al. protocol by using a fully-connected graph.



We implemented the Bell et al. protocol in \system by re-using large
portions of our implementation of the Bonawitz et al. protocol. Our
implementation also generates each client's set of neighbors, and uses
the set of neighbors for each client to determine the construction of
the masks.
The per-client communication cost of our implementation is equivalent
to that of our Bonawitz et al. implementation, but it requires only
$O(k)$ Shamir shares for the $s_{u, v}$ values, achieving the desired
$O(\log n + l)$ complexity.

\subsection{Secret Sharing Sharing Protocol}
\label{sec:sharingsharing}

The Sharing Sharing protocol~\cite{stevens2022secret} builds on the Stevens et al. protocol by using a \emph{two-level} secret sharing scheme. First, each client creates two additive shares of their input (the ``high-level shares''); then, the clients split into two sets of groups of size $\log(n)$ and aggregate their high-level shares by running an aggregation protocol within the small groups. The clients reveal their group-level sums to the server, which cannot reconstruct any single input because of the higher level of additive sharing.
This idea is combined with the LWE-based masking approach of the Stevens et al. protocol to reduce dimensionality, so that the combined protocol scales well with both the number of clients and the size of the input vectors.

We re-used large portions of our implementation of the Stevens et al. protocol to implement the sharing sharing protocol. The main differences are in the high-level sharing and group assignment for clients. The original paper specifies that clients themselves perform reconstruction of the group-level results; we offload this computation to the server instead, under the assumption that the clients will have limited computation power. We reduce the number of server-side reconstructions by adding shares together \emph{before} reconstruction whenever possible.

\subsection{ACORN Protocol}
\label{sec:acorn}

The ACORN protocol~\cite{bell2022acorn} combines the sparser communication graph of the Bell et al. protocol with a key-homomorphic approach for encoding both personal and pairwise masks based on LWE. As in the Stevens et al. protocol, this approach improves computation cost. Unlike Stevens et al., the ACORN protocol generates (homomorphic) encryption keys from public-key agreement, which reduces the dimensionality of the secret-shared data in the protocol. The ACORN protocol is part of a larger contribution that also includes input validation via zero-knowledge proofs.

Our implementation of ACORN leverages the structure of the Bell et al. protocol's implementation, and also re-uses some of the LWE-related implementation from the Stevens et al. protocol. We do not implement the input validation approach proposed in the paper. Rather than implement the paper's proposed packing scheme for plaintexts, we adjust the size of the finite field so that packing for plaintexts is not needed (since it lacks the input validation step, our implementation is not tied to the large finite field required for Bulletproofs).


\section{Evaluation}
\label{sec:evaluation}


In this section, we evaluate the concrete performance of the different secure aggregation protocols in terms of computation time, communication costs, scalability, and the effects of various parameters, such as the number of clients, input vector size, latency, etc. We designed our experiments specifically to answer the following questions:

\begin{itemize}[leftmargin=12pt, itemsep=4pt]
\item \textbf{RQ1} How well do the case study protocols scale with the size of the input vectors?
\item \textbf{RQ2} How well do the case study protocols scale with the number of clients?
\item \textbf{RQ3} What is the effect of network latency on protocol performance?
\item \textbf{RQ4} What is the effect of network bandwidth limits on protocol performance?
\item \textbf{RQ5} What is the overhead of malicious security?
\item \newtext{\textbf{RQ6} Does \system's simulator yield accurate results when compared to actual ``ground truth'' execution?}
\end{itemize}
%

\subsection{Experiment Setup}
Our experiments use \system to evaluate the concrete performance of the protocols described in Section~\ref{sec:case-studies}. Our open-source release contains the code for both the \system framework and the implementations of these protocols. \newtext{
We ran each experiment on a single machine with 32GB of memory, utilizing a high-performance cluster to execute multiple experiments simultaneously.}


We split our experiments into two settings, following two common use cases.
In the ``Large Vectors'' setting, we fixed the number of clients (to 64) and varied the size of the vectors being aggregated.
%
In the ``Few Clients'' setting, we fixed the vector size to 100 and varied the number of clients from 8 to 128. In the ``Many Clients'' setting, we fixed the vector size to 100 and varied the number of clients from 100 to 10,000, in order to test the support of the Bell, ACORN, and Sharing Sharing protocols for a large number of clients.
The input vectors for each protocol do not affect performance, so we used a constant input vector to allow for verifying the correctness of the output. We used a finite field of size $2^{31}-1$. We report the average results and standard error over five runs.

\begin{table}
\centering
\renewcommand{\arraystretch}{1.5}
\begin{tabular}{|p{0.07\textwidth}| p{0.08\textwidth}|p{0.08\textwidth}| p{0.1\textwidth}|}
\hline 
\textbf{Setting} & \textbf{Protocols} & \textbf{Clients} & \textbf{Dimensions}\\
\hline 
{Large Vectors} & All & 64 & $1e^{[1,2,3,4,5]}$ \\
\hline 
{Few Clients} &  All& $2^{[3,4,5,6,7]}$ & 100 \\
\hline 
{Many Clients} & Bell, Sharing sharing, ACORN & [100, 1000, 3000, 5000, 10000] & 100 \\
\hline 
\end{tabular}
\caption{Experiment Settings.}
\label{tbl:experiment_settings}
\end{table}

In all of our experiments, we used the most favorable settings for each protocol that would ensure security in the semi-honest setting. Following Bell et al.~\cite{bell2020secure}, we set both the fraction of malicious clients and the fraction of dropouts to 5\% of the total.
%
For the Stevens et al. protocol, following~\cite{stevens2022efficient}, we set the size of the secret vector $S$ to 710. For the Bell et al. and ACORN protocols, following the most favorable settings of~\cite{bell2020secure}, we set $k=50$. For the sharing sharing protocol, we optimize group according to expected dropouts per group.

\newtext{The Sharing Sharing protocol is not designed for situations with a few clients, and finding the optimal group size for best performance in these smaller settings is challenging. Hence, in our study, we have left out the results of this protocol for scenarios involving a small number of clients,  focusing on its stronger performance in larger settings.}

\subsection{Accuracy of the simulation}
\newtext{Evaluating the accuracy of \system's simulator is difficult by definition---\system is designed to simulate evaluation scenarios that are typically not feasible on a large scale by other means. To address this and validate the reliability of \system, we performed a physical deployment with a smaller number of clients, enabling us to obtain ground truth results. The primary use we envision for \system is comparing different protocols or protocol configurations, either to demonstrate an advance in efficiency or in preparation for deployment of a protocol; for such comparisons, consistency and comparability of results is more important than absolute accuracy of running time. In addition, most of the key components for evaluation (e.g. computation time and size of transmitted messages) are measured directly by the simulator, so these results will be exactly comparable to protocol deployments on equivalent hardware.}

\newtext{In validating \system, we constructed a real network backend using the same client and server classes as those implemented in the simulator. These experiments are conducted within an advanced high-performance computing environment, which includes a SLURM job scheduling system with each node allocated 32 GB of memory. The experiments involve launching a server process and multiple client processes (varying in number and dimensions) across different nodes, thus enabling us to replicate and test our network protocols in a realistic environment. We run our experiments on the Acorn, Baseline, Bell, Bonawitz, Shamir Sharing, and Stevens protocols with 5, 10, and 20 clients, and dimensions ranging from 10 to 100,000.  This approach enables a direct comparison of total running time, providing valuable insights into the performance of the protocols under study in both simulated and real-world network environments.}

\begin{figure*}
\centering
\begin{tabular}{c}
  \hline
  \multicolumn{1}{c}{\textbf{Number of Clients: 5}}\\
  \hline
  \includegraphics[width=1\textwidth]{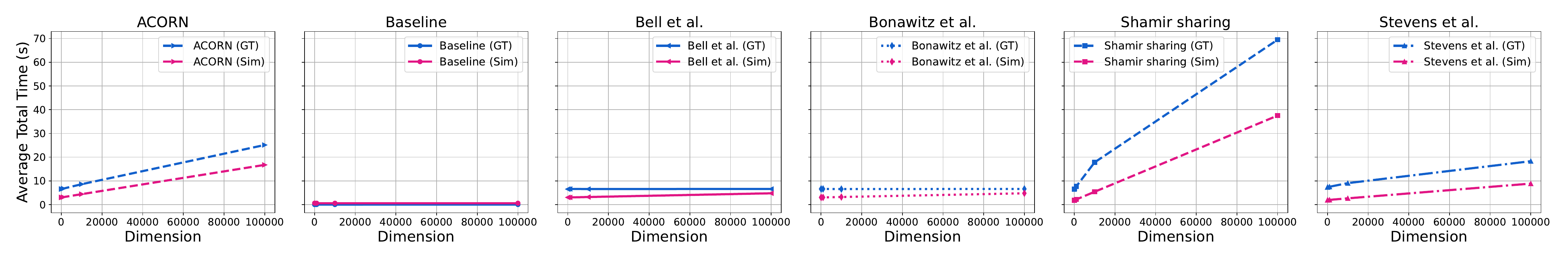}\\
  \hline
  \multicolumn{1}{c}{\textbf{Number of Clients: 10}}\\
  \hline
  \includegraphics[width=1\textwidth]{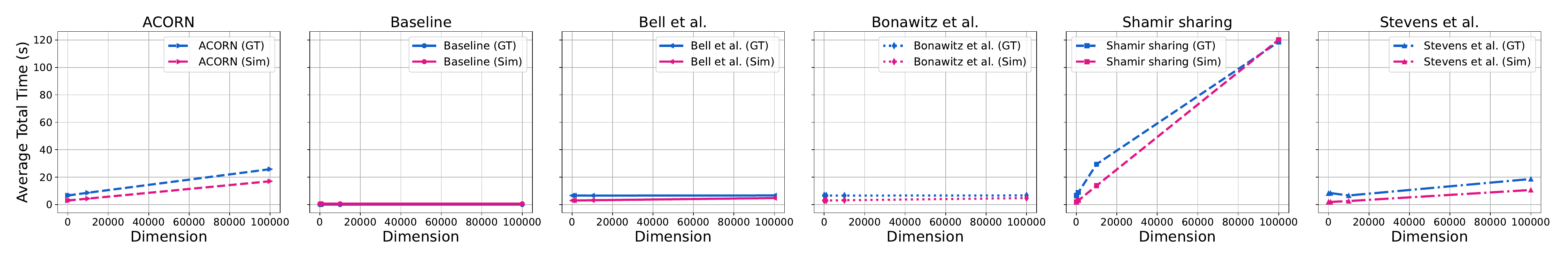}\\
  \hline
  \multicolumn{1}{c}{\textbf{Number of Clients: 20}}\\
  \hline
  \includegraphics[width=1\textwidth]{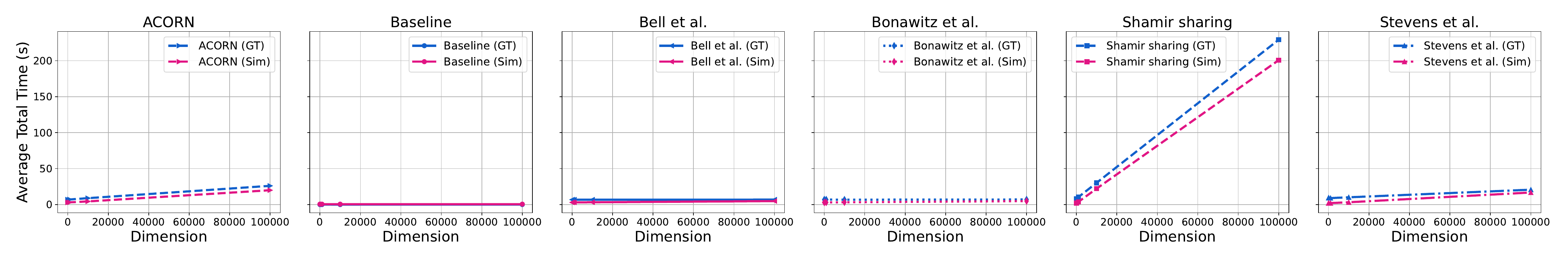}
\end{tabular}
\caption{\newtext{Total running time comparison between simulated and real experiments.}}
\label{fig:groundtruth_results}
\end{figure*}

\newtext{Figure \ref{fig:groundtruth_results} reveals that when comparing simulated results to ground truth results, there is a strong correlation in the performance trends of the various protocols. The simulator exhibits a slight tendency to underestimate total running times, especially in scenarios with fewer clients and smaller dimensions. However, it consistently captures the essential performance and scalability trends of these protocols across different settings. Importantly, this correlation becomes more pronounced as the number of clients increases, with the simulation results for protocols aligning more closely with the ground truth. This trend confirms the simulator's ability to model complex network behaviors accurately, showcasing \system's effectiveness in simulating diverse protocol behaviors.}

\subsection{Results: Semi-Honest Security}

This section describes our comparison between semi-honest variants of the case study protocols. 


\begin{figure*}
\centering
\begin{tabular}{c c c}
  \hline
  \multicolumn{3}{c}{\textbf{Total Running Time (Semi-Honest)}}\\
  \hline
  \textbf{\footnotesize (a) Large Vectors (64 clients)} 
& \textbf{\footnotesize (b) Few Clients (length-100 vectors)} 
& \textbf{\footnotesize (c) Many Clients (length-100 vectors)}\\
\mkgraph{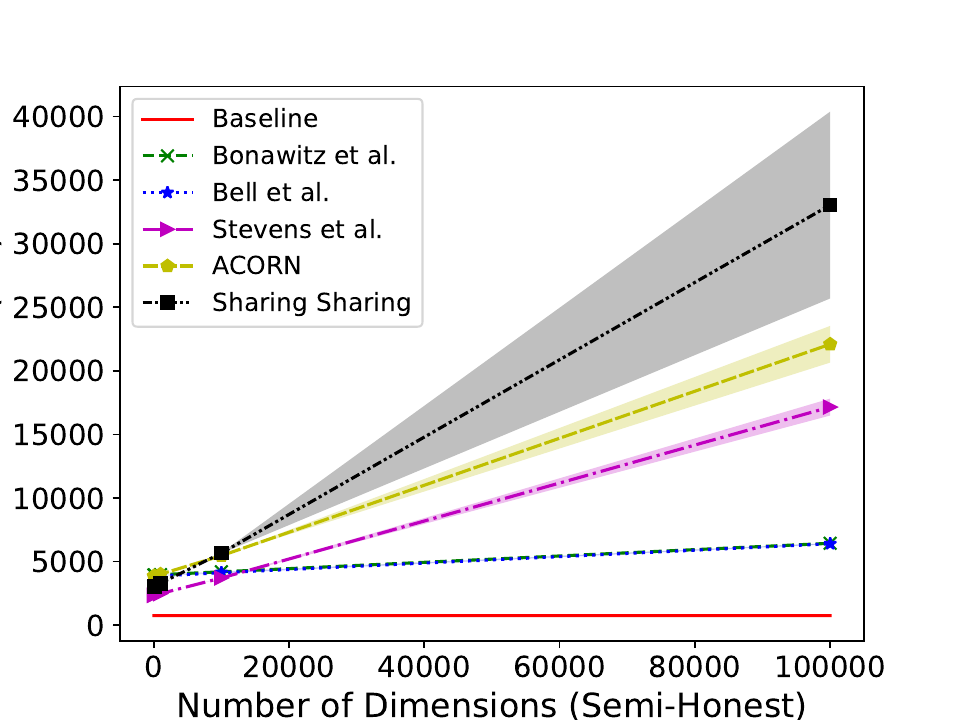}
& \mkgraph{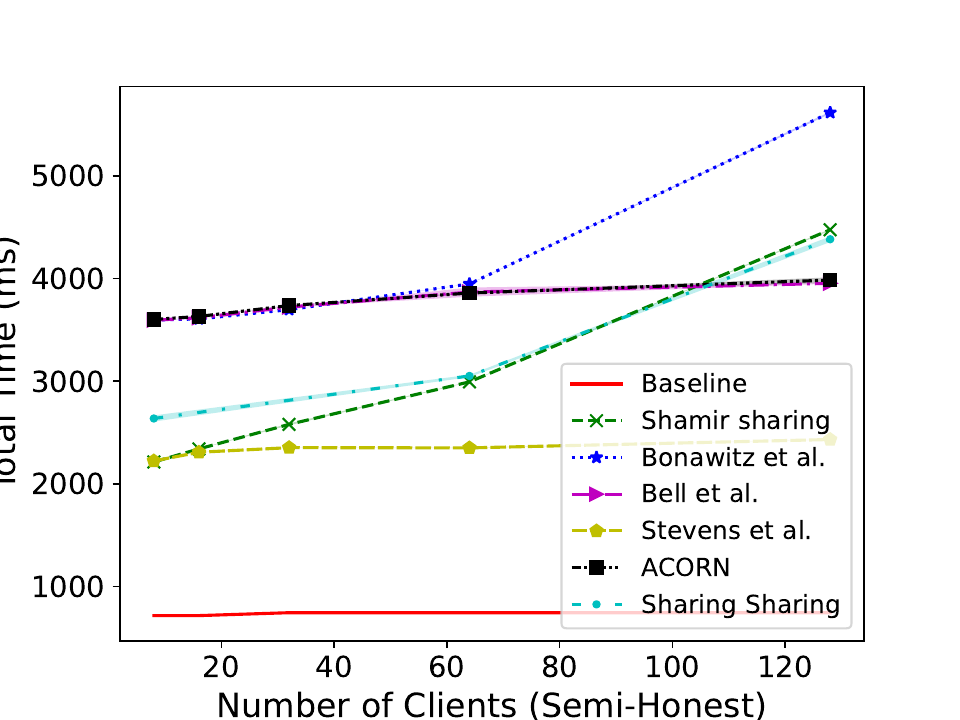}
& \mkgraph{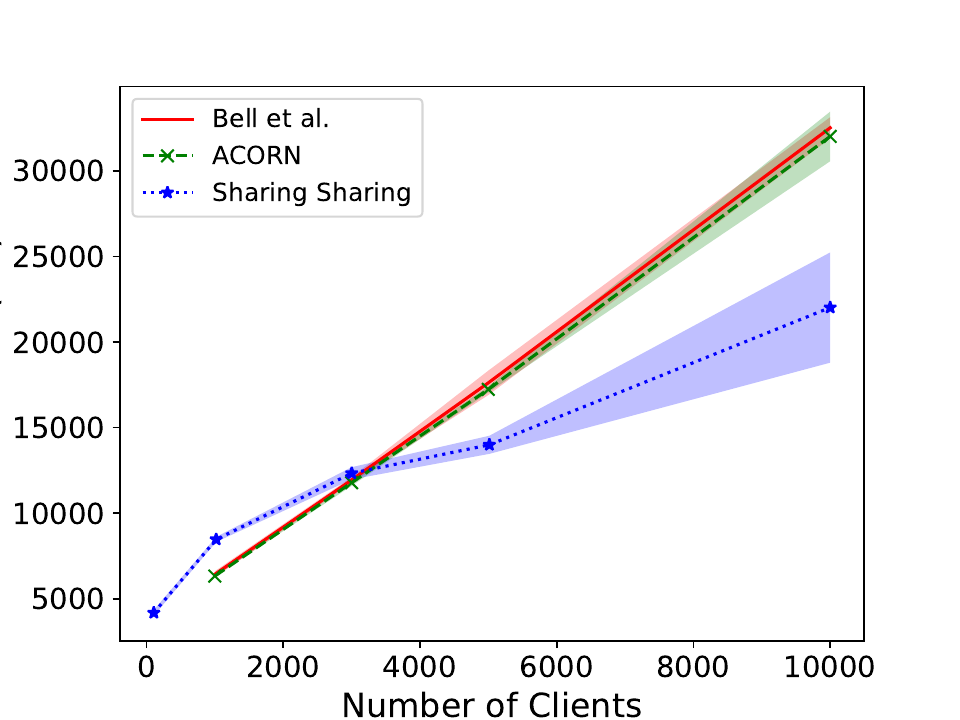}\\
  \hline
\end{tabular}
\caption{Total running time comparison between semi-honest protocols. Shaded area indicates standard error.}
\label{fig:results1}
\end{figure*}

\paragraph{Total time.}
Figure~\ref{fig:results1} summarizes the total running time of the case study protocols in the three experimental settings. In the Large Vector setting (Figure~\ref{fig:results1}(a)), all protocols scale well with the dimensionality of the aggregated vectors; the Bell et al. and Bonawitz et al. protocols yield the lowest total running times. The others require matrix operations (by both the client and the server) that contribute to increasing computation time as vector size increases. In the Few Clients setting (Figure~\ref{fig:results1}(b)), the Stevens et al. is slightly better than the Bell et al. and Bonawitz et al. protocols, due to the smaller vectors. In the Many Clients setting (Figure~\ref{fig:results1}(c)), all three protocols scale well with the number of clients; in contrast to the other settings, the Sharing Sharing protocol yields the best concrete performance when the number of clients exceeds 3000.

\begin{figure*}
\centering
\begin{tabular}{c c c}
  \hline
  \multicolumn{3}{c}{\textbf{Server Computation (Semi-Honest)}}\\
  \hline
\textbf{\footnotesize (a) Large Vectors (64 clients)} 
& \textbf{\footnotesize (b) Few Clients (length-100 vectors)} 
& \textbf{\footnotesize (c) Many Clients (length-100 vectors)}\\
\mkgraph{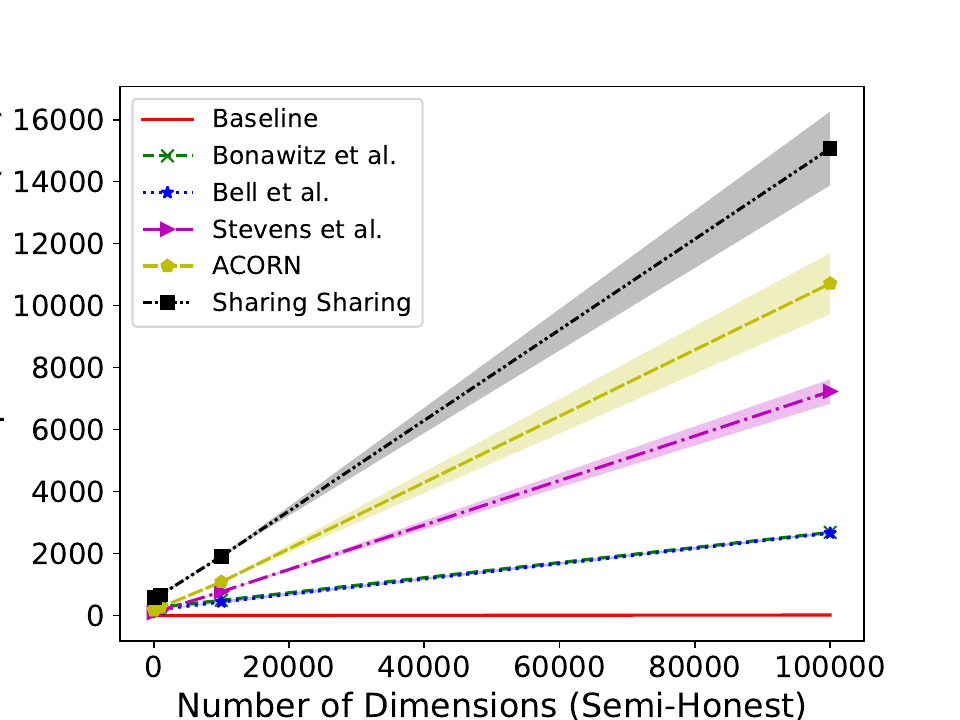}
& \mkgraph{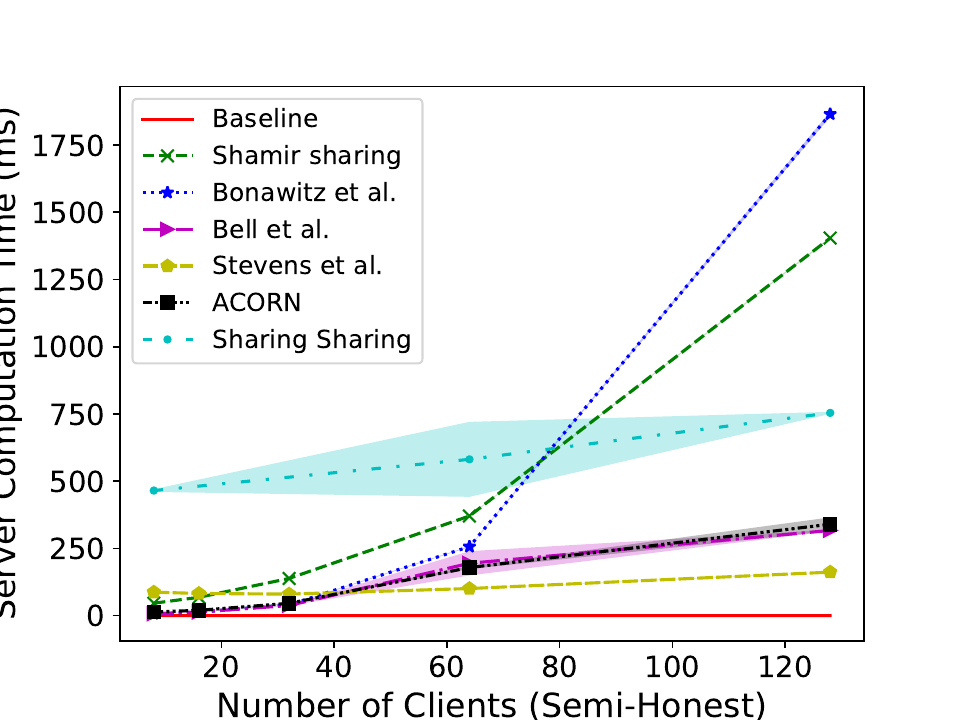}
& \mkgraph{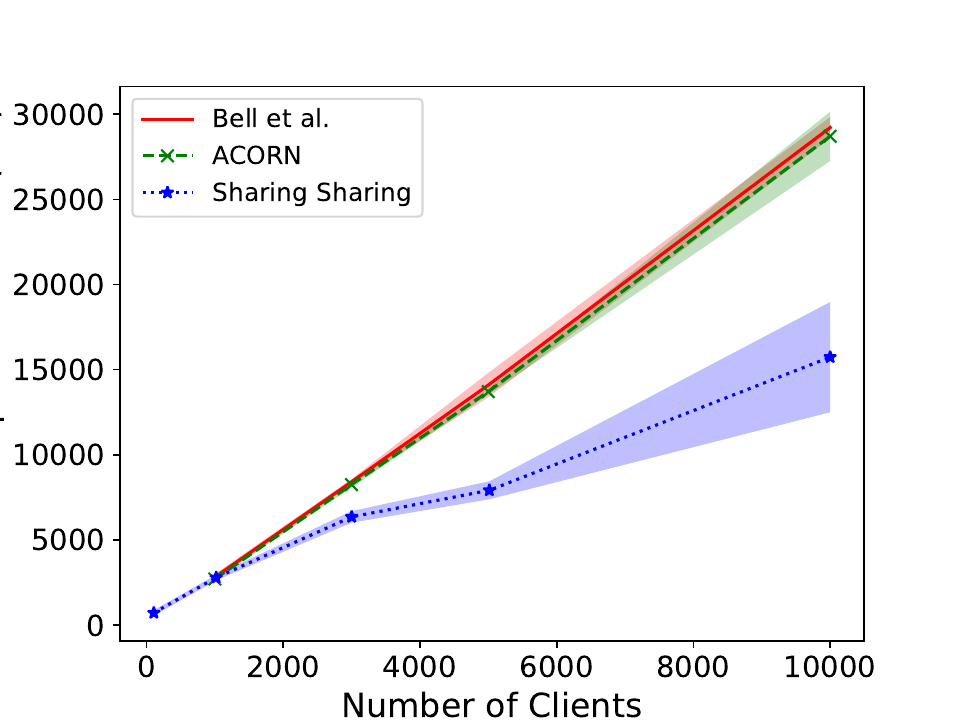}\\
  \hline
  \multicolumn{3}{c}{\textbf{Avg Client Computation (Semi-Honest)}}\\
  \hline
\textbf{\footnotesize (a) Large Vectors (64 clients)} 
& \textbf{\footnotesize (b) Few Clients (length-100 vectors)} 
& \textbf{\footnotesize (c) Many Clients (length-100 vectors)}\\
\mkgraph{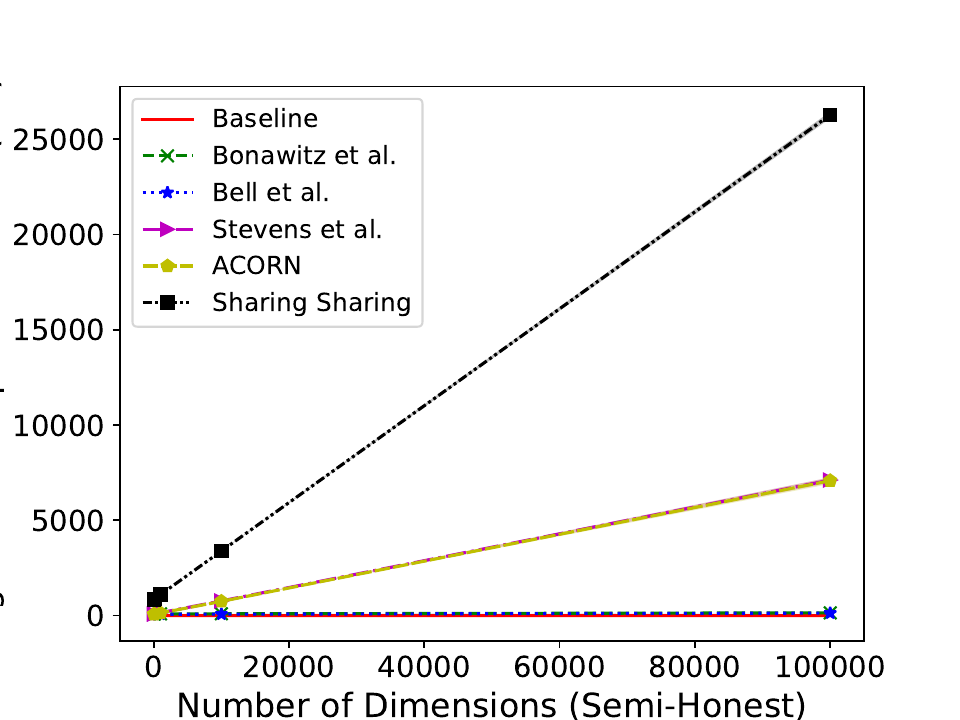}
& \mkgraph{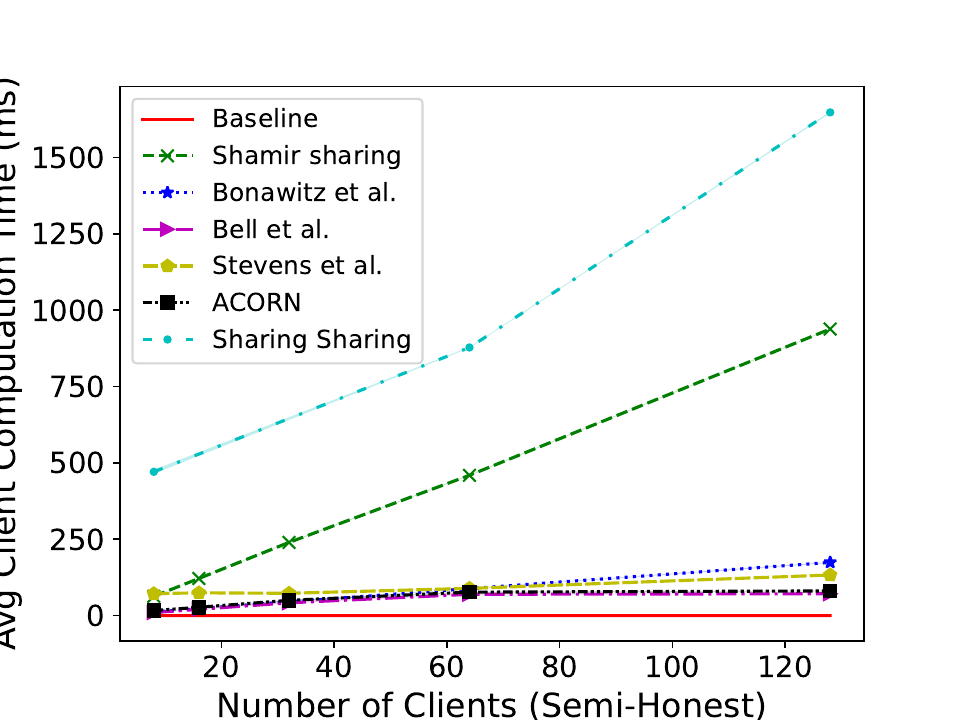}
& \mkgraph{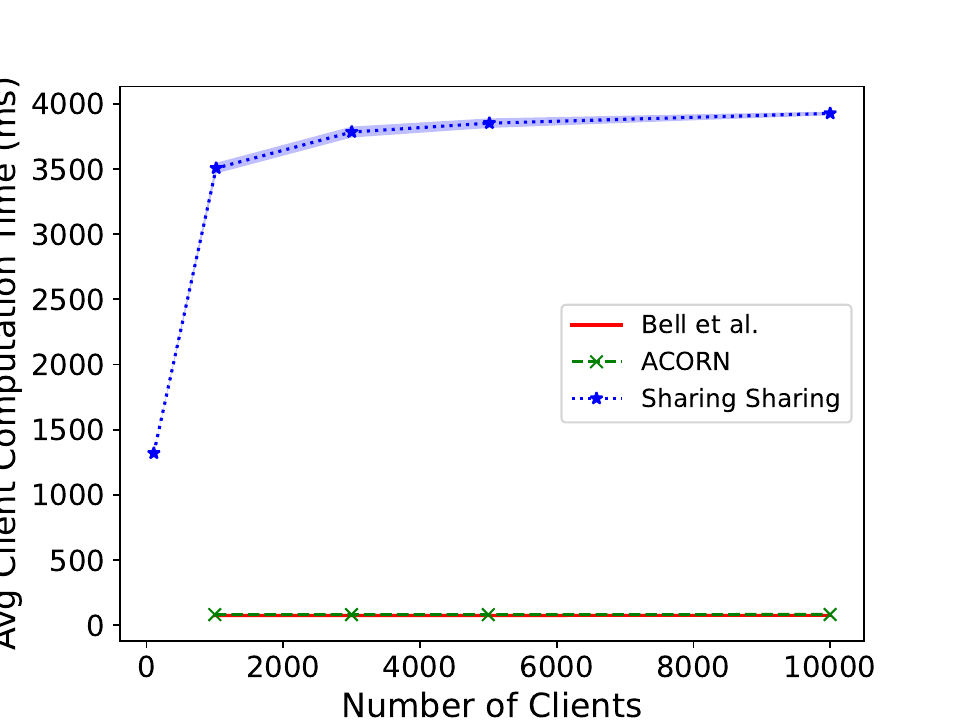}\\
\hline
\end{tabular}
\caption{Computation time comparison between semi-honest protocols. Shaded area indicates standard error.}

\label{fig:results2}
\end{figure*}

\paragraph{Computation time.}
Figure~\ref{fig:results2} summarizes the server and client computation time for each protocol in each of the three settings. Comparing these results to Figure~\ref{fig:results1}, it is clear that computation time is the most important factor in overall running time. In the Large Vectors setting (Figure~\ref{fig:results2}(a)), server computation time is largest for the Stevens et al., ACORN, and Sharing Sharing protocols (due to the combination of matrix operations and reconstructions required on the server); client computation time is largest for the LWE-based protocols due to matrix operations. The same trend continues in the other settings---in the Few Clients setting (Figure~\ref{fig:results2}(b)), the Bell et al. and Bonawitz et al. protocols perform best with server computation times for clients under 40, but for larger client counts, the Stevens et al. protocol becomes more efficient.
In the Many Clients setting (Figure~\ref{fig:results2}(c)), the Bell et al. and ACORN protocols have the lowest client computation time, but the Sharing Sharing protocol has a much lower server computation time, which results in the low total time seen earlier.

\begin{figure}
  \centering
\vspace*{-10pt}\hspace*{-10pt} \includegraphics[width=.5\textwidth]{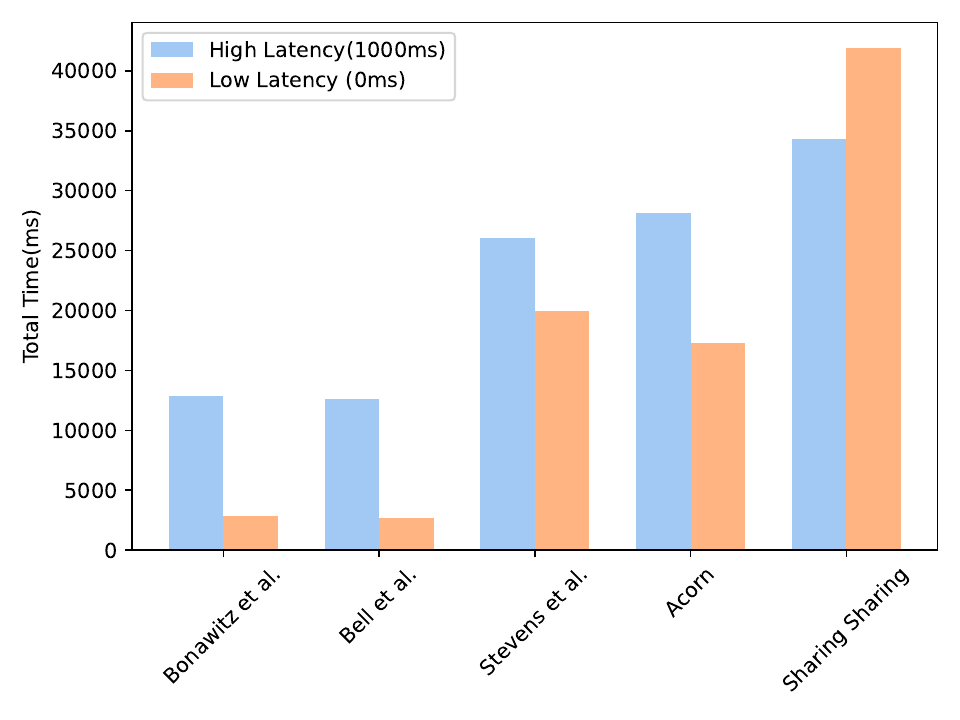}
  \caption{Effect of latency on total running time (semi-honest security). Latency impacts all protocols in roughly the same way, since all protocols have similar round complexity.}
  \label{fig:results3}
\end{figure}

\paragraph{Network latency.}
Our previous experiments use \system's model of network latency derived from actual internet speed test data. To answer RQ3, we varied the latency model to evaluate the impact of latency on total running time. Figure~\ref{fig:results3} presents the results. All of our case study protocols have fairly low round complexity, so even a huge increase in latency (from 0ms to 1000ms per message) does not cause a large increase in total running time. These results suggest that the case study protocols are likely to perform well even over high-latency WAN connections.

\begin{figure}
  \centering
  \vspace*{-10pt} \hspace*{-10pt} \includegraphics[width=.5\textwidth]{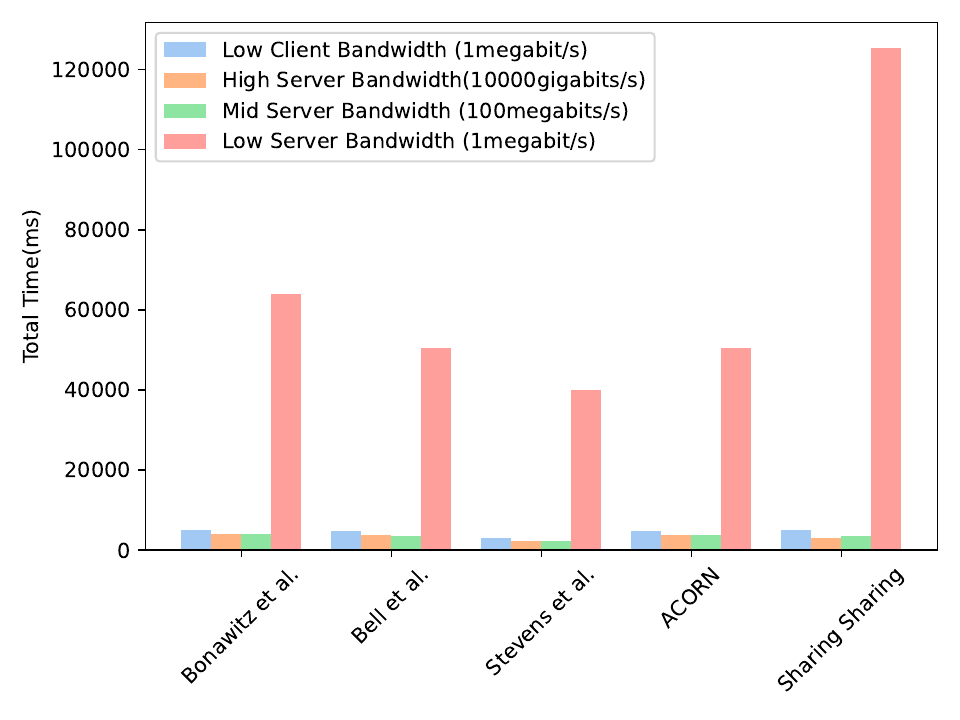}
  \caption{Effect of bandwidth limitations on total running time (semi-honest security), with 64 clients and length-100 vectors. Limits on client bandwidth have a small impact, but limits on server bandwidth have a large impact.}
  \label{fig:results4}
\end{figure}

\paragraph{Network bandwidth limitations.}
Our previous experiments assume that bandwidth is unlimited, both for the server and for the clients. To evaluate the importance of bandwidth limitations for protocol performance, we evaluated each protocol with a 1mbps limit on client bandwidth and with a 1mbps limit on server bandwidth. The results appear in Figure~\ref{fig:results4}. Limiting client bandwidth has little impact on total running time, but limiting server bandwidth has a large effect on all protocols.

\begin{figure*}
\centering
\begin{tabular}{c c c}
  \hline
  \multicolumn{3}{c}{\textbf{Server Bytes Received (Semi-Honest)}}\\
  \hline
  \textbf{\footnotesize (a) Large Vectors (64 clients)} 
& \textbf{\footnotesize (b) Few Clients (length-100 vectors)} 
& \textbf{\footnotesize (c) Many Clients (length-100 vectors)}\\
\mkgraph{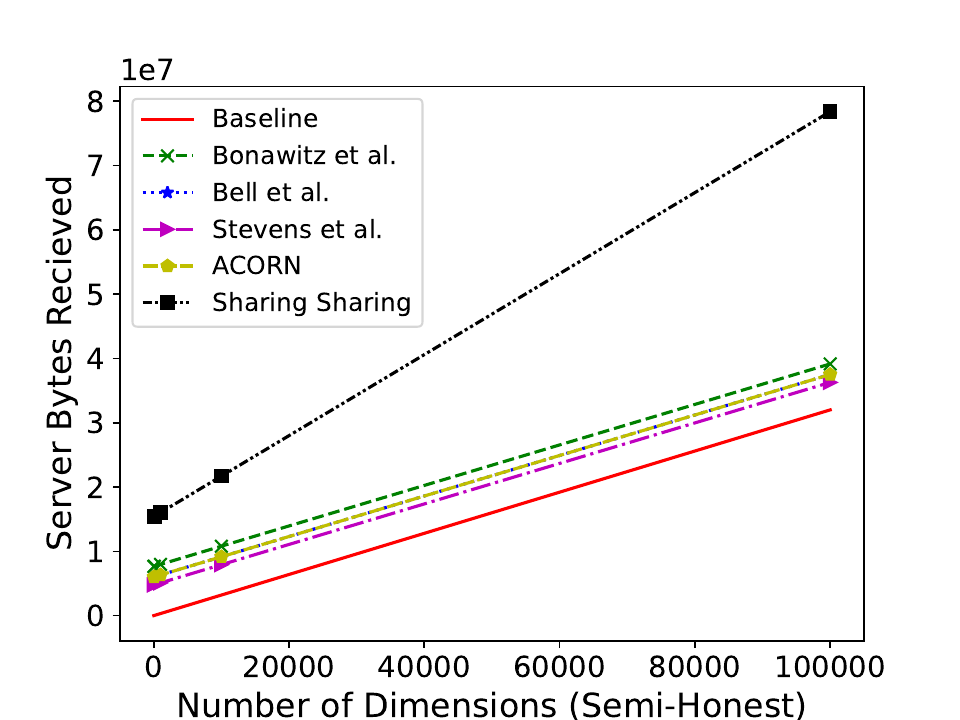}
& \mkgraph{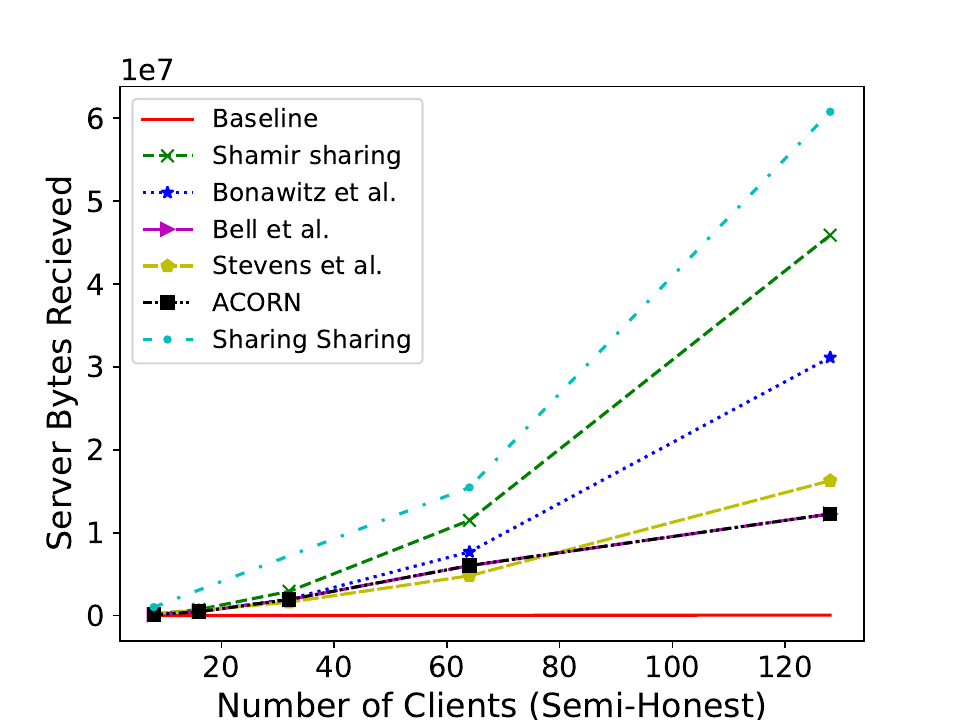}
& \mkgraph{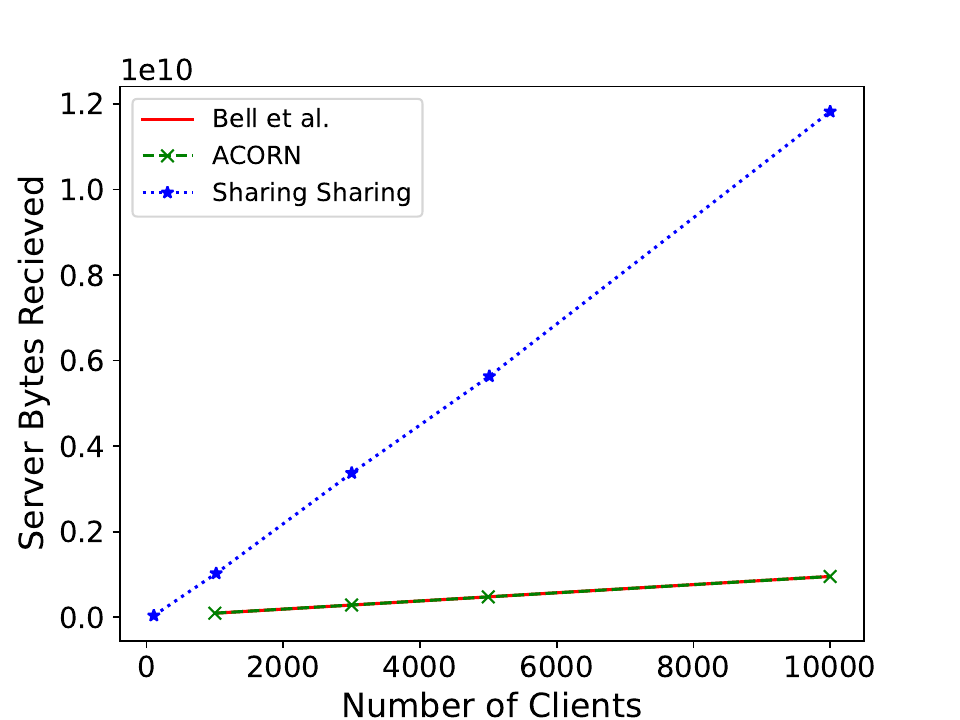}\\
  \hline
\end{tabular}
\caption{Server bytes received comparison between semi-honest protocols. }
\label{fig:results5}
\end{figure*}

This effect is directly linked to the amount of traffic the server receives. The total bytes received by the server appear in Figure~\ref{fig:results5}. The amount of traffic received by the server increases more quickly with vector size for the Sharing Sharing protocol than the others (Figure~\ref{fig:results5}(a), (b), and (c)). Complete results for communication cost, including the traffic sent and received by clients (which mirrors the traffic received by the server) appear in Appendix~\ref{sec:appendix_results}.

\subsection{Results: Malicious Security}

\begin{figure*}
\centering
\begin{tabular}{c c}
  \hline
  \multicolumn{2}{c}{\textbf{Total Running Time (Malicious-Secure)}}\\
  \hline
  \textbf{\footnotesize (a) Large Vectors (64 clients)} 
& \textbf{\footnotesize (b) Few Clients (length-100 vectors)} \\
\mkgraph{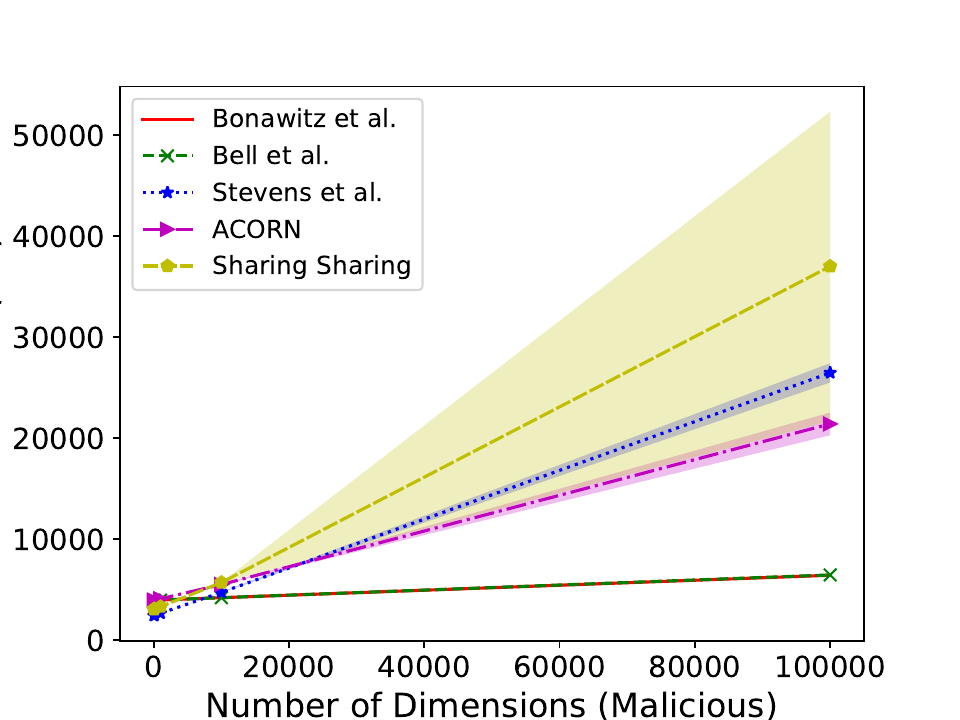}
& \mkgraph{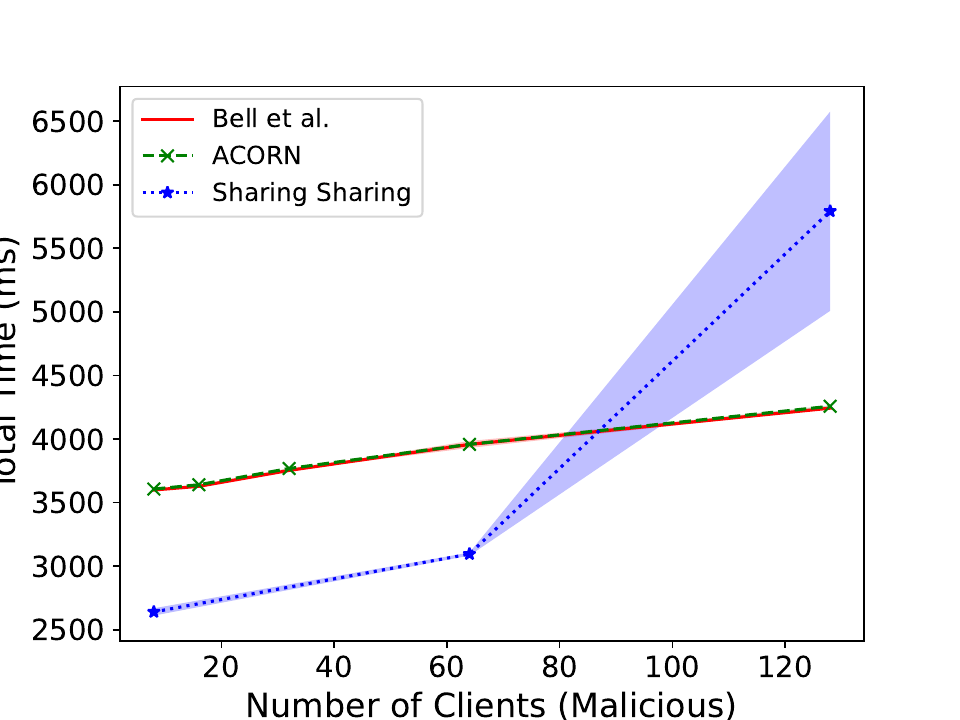}\\
  \hline
\end{tabular}
\caption{Total running time comparison between malicious-secure protocols. Shaded area indicates standard error.}
\label{fig:results6}
\end{figure*}

\begin{table}
  \centering
  \begin{tabular}{l l l l}
    \hline
    \textbf{Protocol} & \textbf{Semi-honest} & \textbf{Malicious} & \textbf{Overhead}\\
    \hline
    Bonawitz et al. & 2938ms & 3795ms & 29\% \\
    Bell et al. & 3879ms & 3938ms & 1.5\%\\
    Stevens et al. & 4104ms & 5930ms & 44\% \\
    ACORN & 5649 & 5796ms & 2.6\%\\
    Sharing Sharing & 10269ms & 23572ms & 129\%\\
    \hline
  \end{tabular}
  \caption{Precise results comparing semi-honest and malicious-secure variants of protocols, for 64 clients and length-10,000 vectors.}
  \label{tbl:results7}
\end{table}
    
To evaluate the cost of malicious security, we re-ran our experiments with the malicious-secure variants of all our case study protocols. The total running time results appear in Figure~\ref{fig:results6}. The graphs are similar to the corresponding results in the semi-honest case, supporting the claim that the malicious-secure variants of these protocols incur only modest overhead. More precise results appear in Table~\ref{tbl:results7}, for the case of 64 clients and length-10,000 vectors. Both sets of results suggest that the overhead of malicious security is reasonable for all protocols, and especially good for the Bell et al. and ACORN protocols. The Sharing Sharing protocol performs poorly in this setting. Additional results for malicious-secure protocols appear in Appendix~\ref{sec:appendix_results2}.

\subsection{Discussion}


\paragraph{Protocol comparison.}
Our results show that all the protocols have the expected asymptotic
performance, but they differ significantly in terms of concrete
performance. For a small number of clients (\textbf{RQ1}), the Bonawitz et al. and Bell et al. protocols provide the lowest total time. For a
very large number of clients (\textbf{RQ2}), the Sharing Sharing protocol provides the best performance; the Bell et al. protocol may outperform Sharing Sharing for larger vectors.




\paragraph{Computation vs. communication.}
In all settings, the results suggest that computation time is the largest contributing factor to total running time of the protocol (\textbf{RQ3}). All of our case study protocols involve a small constant number of communication rounds and attempt to minimize total communication cost. Secure aggregation protocols with similar properties are likely to follow the same pattern.

\paragraph{Computation and constant factors.}
However, our results suggest that different protocols can have very different computation costs, and can distribute those costs between the clients and the server differently. For example, LWE-based approaches like Stevens et al., ACORN, and Sharing Sharing have higher computational costs for clients than the Bonawitz et al. and Bell et al. protocols, but tend to have lower computational costs for the server when vectors are small. The Sharing Sharing protocol in particular has much higher computation cost for the client, but much lower cost for the server in the Many Clients setting.
For large vectors, however, the computation cost of the matrix operations required in these approaches means that the Bell et al. and Bonawitz et al. approaches often provide better performance.
%
These results point to the importance of end-to-end evaluation of protocols in determining overall performance---such effects are typically impossible to completely understand only by analyzing the asymptotic behaviors of protocols, and depend on the deployment scenario.

\paragraph{Importance of latency (RQ3).}
The results of our network latency experiment suggest that typical WAN latencies are unlikely to have a major impact on the total running time of our case study protocols (\textbf{RQ3}). This result suggests that protocol design should focus on concrete performance improvements in computation time, rather than on further reducing round complexity.

\paragraph{Importance of bandwidth (RQ4).}
The results of our network bandwidth limitation experiment suggest that limits on client bandwidth are not likely to have significant impact on the running time of the case study protocols, but that limits on server bandwidth can have a huge effect (\textbf{RQ4}). Protocols differ slightly in the impact of server bandwidth limitations, but all are affected. These results suggest that deployments of all of these protocols must include adequate server bandwidth to ensure good performance.

\paragraph{Overhead of malicious security (RQ5).}
Our comparison between semi-honest and malicious variants of the case study protocols shows that the overhead of malicious security is below 50\% for all protocols, and below 5\% for the Bell et al. and ACORN protocols. These results suggest that malicious security can be achieved with good performance.

\paragraph{Feasibility of end-to-end evaluation (RQ6).}
Our evaluation results demonstrate the feasibility of end-to-end empirical evaluation of protocols with \system, even at scales well beyond the feasibility of traditional evaluation. Our experiments involved up to 10,000 clients and vectors with millions of elements; these experiments ran on a single machine with 128GB of memory in just a few hours. \newtext{Our validation experiment shows \system's simulator accurately models the performance and scalability properties of protocols, and that its accuracy improves with the number of clients involved with the protocol---probably because the impact of execution overhead (e.g. constructing and managing sockets and serializing messages) is smaller as the number of clients increases.}

  

\section{Related Work}

\paragraph{Secure aggregation.}
As described earlier, \emph{secure aggregation} protocols are lightweight multiparty computation protocols specifically designed for summing up large vectors, and are primarily designed to facilitate federated learning applications.

The first practical protocol for the large-vector setting was due to Bonawitz et al.~\cite{bonawitz2017practical}, while the first protocol for the many-client setting was due to Bell et al.~\cite{bell2020secure}. Since then, several other protocols have been developed that make additional improvements.
Among these, only Turbo-Aggregate~\cite{so2021turbo} attempts to improve performance in the many-client setting; it reconstructs masks among subsets of users (rather than pairwise), but has a weaker threat model than the Bell et al.~\cite{bell2020secure} protocol.

In the large-vector setting, several new protocols have been recently proposed, including  MicroFedML~\cite{guo2022microfedml} (reduces round complexity for sparse gradients), LightSecAgg~\cite{yang2021lightsecagg} and the protocol of Stevens et al.~\cite{stevens2022efficient} (improve concrete performance for dropouts), FastSecAgg~\cite{kadhe2020fastsecagg} (improves performance using Fast Fourier Transform).
All of these protocols can be simulated in \system, and are important targets for future empirical evaluations.

\paragraph{Input validation.}
Recent work has made progress towards ensuring input validity in secure aggregation. EIFFeL~\cite{roy2022eiffel} requires each client to produce a zero-knowledge proof that their input is within a reasonable range; ACORN~\cite{bell2022acorn} improves on the performance of EIFFeL by leveraging more efficient aggregation of these proofs. Protocols that integrate input validation have even more complicated concrete performance properties and are an exciting future target for evaluation with \system.

\paragraph{Secure machine learning.}
Many approaches have been developed for machine learning \emph{without} secure aggregation~\cite{truex2019hybrid, fereidooni2021safelearn, ryffel2020ariann,
  davidson2021star, jayaraman2021revisiting}. These approaches often ask clients to secret-share model updates between two servers, and run an MPC protocol between the two servers. These approaches have a much weaker threat model than secure-aggregation-based approaches, since they require non-collusion between the two servers. In addition, they do not present the same challenges for empirical evaluation as secure aggregation protocols, since only the two servers need to be simulated; while these approaches could be simulated using \system, a simulation-based approach is not required for empirical evaluation of these approaches.

%


\paragraph{General-purpose MPC.}
General-purpose MPC protocols evaluate arithmetic or boolean circuits, and therefore can (in principle) implement any function~\cite{yao1986generate, bgw, gmw, bmr, spdz}. Such protocols are generally designed for two parties (2PC), three parties (3PC), or a small number of parties (MPC)---none are designed for the many-client setting. General-purpose MPC protocols can therefore be empirically evaluated without the use of a simulation framework like \system, though for recent work on larger-scale protocols (e.g. up to 128 parties~\cite{wang2017global} or even tens or hundreds of thousands~\cite{gordon2021more}), \system may make implementation and evaluation simpler.

\paragraph{Other applications of secure aggregation.}
Several systems have been developed for differentially private \emph{analytics} (i.e. database queries) that leverage ideas from secure aggregation, including Honeycrisp~\cite{roth2019honeycrisp}, Orchard~\cite{roth2020orchard}, and Crypt$\epsilon$~\cite{roy2020crypt}. These systems are designed to scale to millions of participants, using specialized protocols and a slightly weaker threat model. Because they are designed for the many-client setting, these systems are also challenging to evaluate empirically; \system may also be useful for evaluation of such systems.

\paragraph{Simulation of distributed protocols.}
For protocols designed for a small number of clients (including most general-purpose MPC protocols), experimental evaluation is feasible using actual hardware. As described earlier, this kind of evaluation is essentially impossible in the many-client setting. The Bonawitz et al. protocol~\cite{bonawitz2017practical} included an experimental evaluation of a Java implementation, which was possible due to the relatively small number of clients. Due to its scale, the later Bell et al. protocol~\cite{bell2020secure} included concrete results only for the number of neighbors each client must communicate with, and did not include an experimental evaluation of computation or communication cost. Similarly, Honeycrisp~\cite{roth2019honeycrisp} and Orchard~\cite{roth2020orchard} include concrete experimental results for parts of their protocol, but do not perform end-to-end experiments due to the protocol's intended scale.

The \abides framework~\cite{byrd2019abides} was originally designed to simulate agents in a financial market at large scale. It has been previously applied in an \emph{ad-hoc} manner to experimentally evaluate few-client secure protocols~\cite{byrd2022collusion, guo2022microfedml, byrd2020differentially}. \system builds on \abides to provide a framework for designing, building, evaluating, and refining secure protocols, with a particular focus on the many-client setting.

\section{Conclusion}

We have presented \system, a framework for designing, building, evaluating, and refining secure aggregation protocols. \system enables experimental evaluation with thousands of clients---a setting in which evaluation on actual hardware is not possible.
The \system framework provides a simulator that accurately measures the end-to-end running time of protocols, and includes a domain-specific language (DSL) embedded in Python for defining synchronous secure aggregation protocols. 
We have used \system to conduct an empirical comparison between several existing protocols implemented in our case studies, yielding new insights about the concrete performance of these protocols.
We release the \system framework and our case study implementations as open source, and hope that \system will be useful as a standardized implementation and evaluation tool for new protocols.

\section*{Acknowledgments}

We thank Timothy Stevens for his help implementing the Stevens et al.
protocol~\cite{stevens2022efficient}. This material is based upon work
supported by the National Science Foundation under Grant No. 2238442 and by DARPA under Contract No. HR001120C0087. Any opinions,
findings and conclusions or recommendations expressed in this material
are those of the author(s) and do not necessarily reflect the views of
the National Science Foundation or DARPA. Computations were performed on the Vermont Advanced Computing Center supported in part by NSF award No. OAC-1827314.



%

\bibliographystyle{plain}
\bibliography{refs}




\appendices

\section{Additional Evaluation Results: Sharing Sharing Protocol}
\label{sec:appendix_results3}

\begin{figure*}
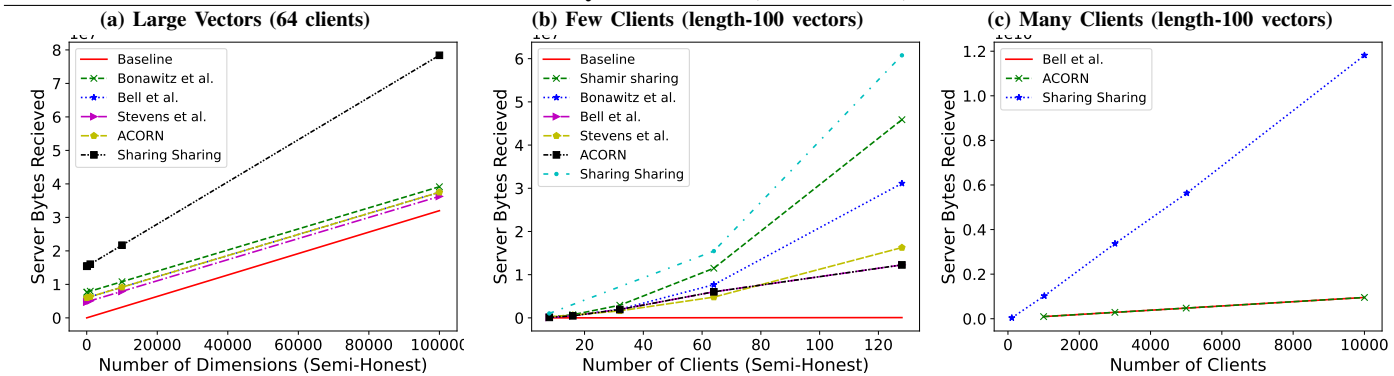

\centering
\begin{tabular}{c c c}
  \hline
  \multicolumn{3}{c}{\textbf{Total Running Time (Semi-Honest)}}\\
  \hline
  \textbf{\footnotesize (a) Large Vectors (64 clients)} 
& \textbf{\footnotesize (b) Few Clients (length-100 vectors)} 
& \textbf{\footnotesize (c) Many Clients (length-100 vectors)}\\
\mkgraph{final_figures/dim_total_time_ms_results.pdf}
& \mkgraph{final_figures/client_total_time_ms_results.pdf}
& \mkgraph{final_figures/large_client_total_time_ms_results.pdf}\\
  \hline
  \multicolumn{3}{c}{\textbf{Server Computation (Semi-Honest)}}\\
  \hline
\textbf{\footnotesize (a) Large Vectors (64 clients)} 
& \textbf{\footnotesize (b) Few Clients (length-100 vectors)} 
& \textbf{\footnotesize (c) Many Clients (length-100 vectors)}\\
\mkgraph{final_figures/dim_server_computation_time_ms_results.pdf}
& \mkgraph{final_figures/client_server_computation_time_ms_results.pdf}
& \mkgraph{final_figures/large_client_server_computation_time_ms_results.pdf}\\
  \hline
  \multicolumn{3}{c}{\textbf{Avg Client Computation (Semi-Honest)}}\\
  \hline
\textbf{\footnotesize (a) Large Vectors (64 clients)} 
& \textbf{\footnotesize (b) Few Clients (length-100 vectors)} 
& \textbf{\footnotesize (c) Many Clients (length-100 vectors)}\\
\mkgraph{final_figures/dim_avg_client_computation_time_ms_results.pdf}
& \mkgraph{final_figures/client_avg_client_computation_time_ms_results.pdf}
& \mkgraph{final_figures/large_client_avg_client_computation_time_ms_results.pdf}\\
\hline
  \multicolumn{3}{c}{\textbf{Server Bytes Received (Semi-Honest)}}\\
  \hline
  \textbf{\footnotesize (a) Large Vectors (64 clients)} 
& \textbf{\footnotesize (b) Few Clients (length-100 vectors)} 
& \textbf{\footnotesize (c) Many Clients (length-100 vectors)}\\
\mkgraph{final_figures/dim_server_bytes_recieved_results.pdf}
& \mkgraph{final_figures/client_server_bytes_recieved_results.pdf}
& \mkgraph{final_figures/large_client_server_bytes_recieved_results.pdf}\\
  \hline
\end{tabular}
\caption{Comparison between semi-honest protocols, including Sharing Sharing. Shaded area indicates standard error.}
\label{fig:results2-2}
\end{figure*}
Figure~\ref{fig:results2-2} contains the full results for all semi-honest protocols, including the Sharing Sharing protocol. This protocol is excluded from the Large Vectors and Few Clients settings in the main body of the paper because it is not competitive, and including it makes the other protocols more difficult to compare.

\section{Additional Evaluation Results: Semi-Honest Security}
\label{sec:appendix_results}

\begin{figure*}
\centering
\begin{tabular}{c c c}
  \hline
  \multicolumn{3}{c}{\textbf{Avg Client Bytes Sent (Semi-Honest)}}\\
  \hline
\textbf{\footnotesize (a) Large Vectors (64 clients)} 
& \textbf{\footnotesize (b) Few Clients (length-100 vectors)} 
& \textbf{\footnotesize (c) Many Clients (length-100 vectors)}\\
\mkgraph{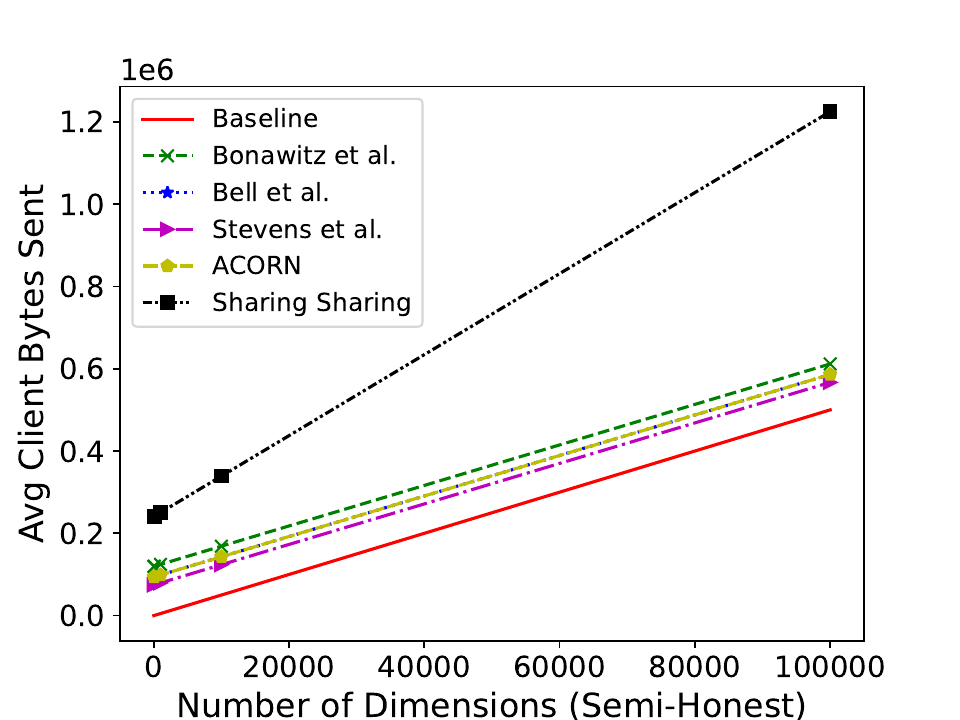}
& \mkgraph{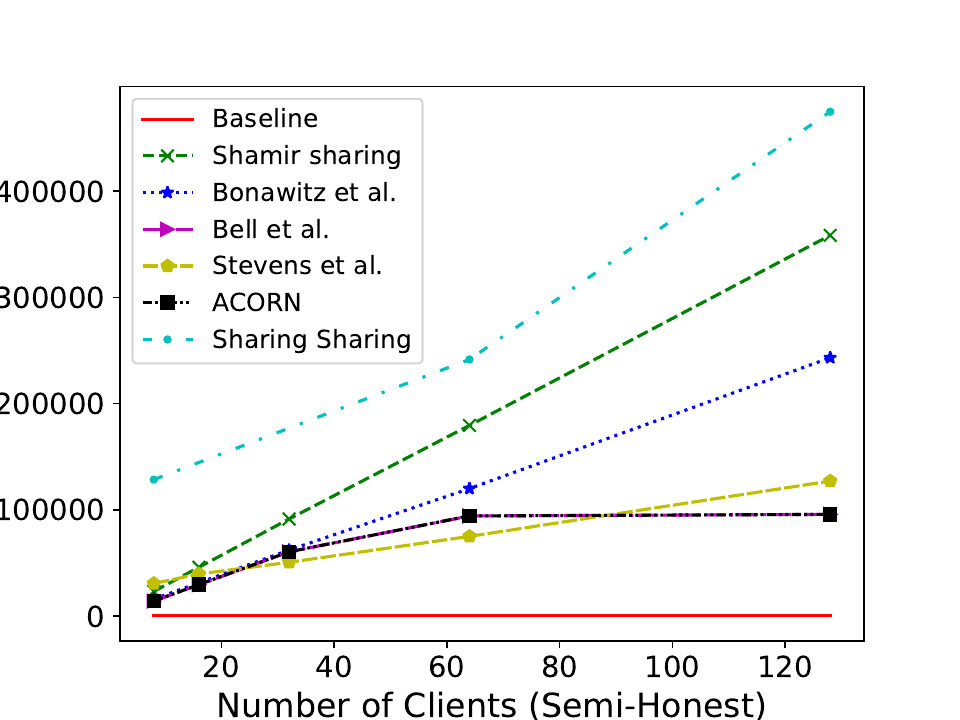}
& \mkgraph{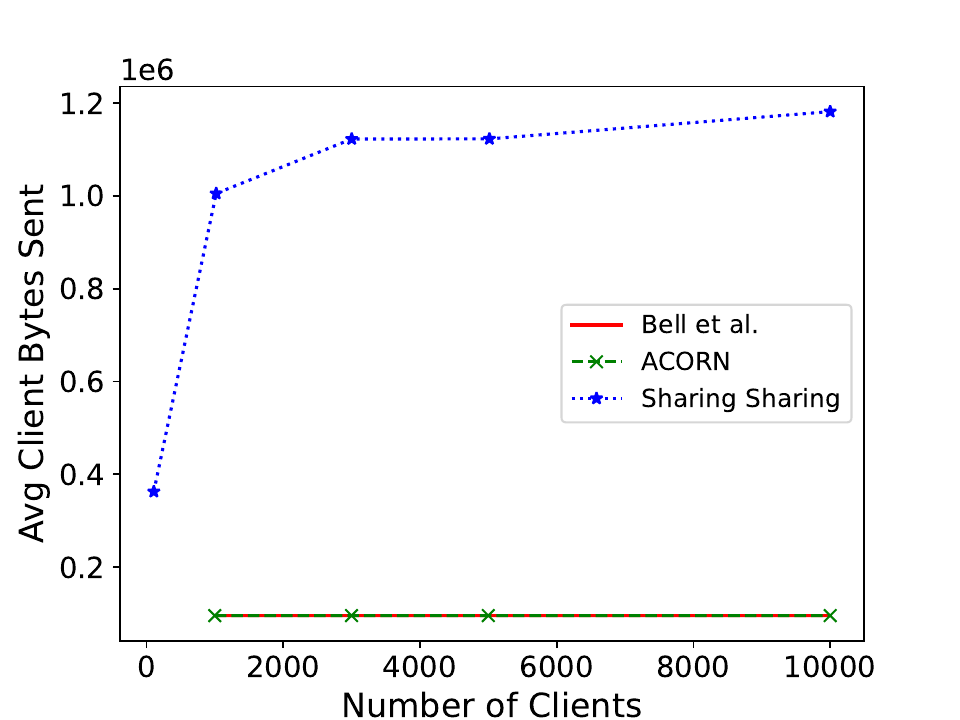}\\
  \hline
  \multicolumn{3}{c}{\textbf{Avg Client Bytes Received (Semi-Honest)}}\\
  \hline
\textbf{\footnotesize (a) Large Vectors (64 clients)} 
& \textbf{\footnotesize (b) Few Clients (length-100 vectors)} 
& \textbf{\footnotesize (c) Many Clients (length-100 vectors)}\\
\mkgraph{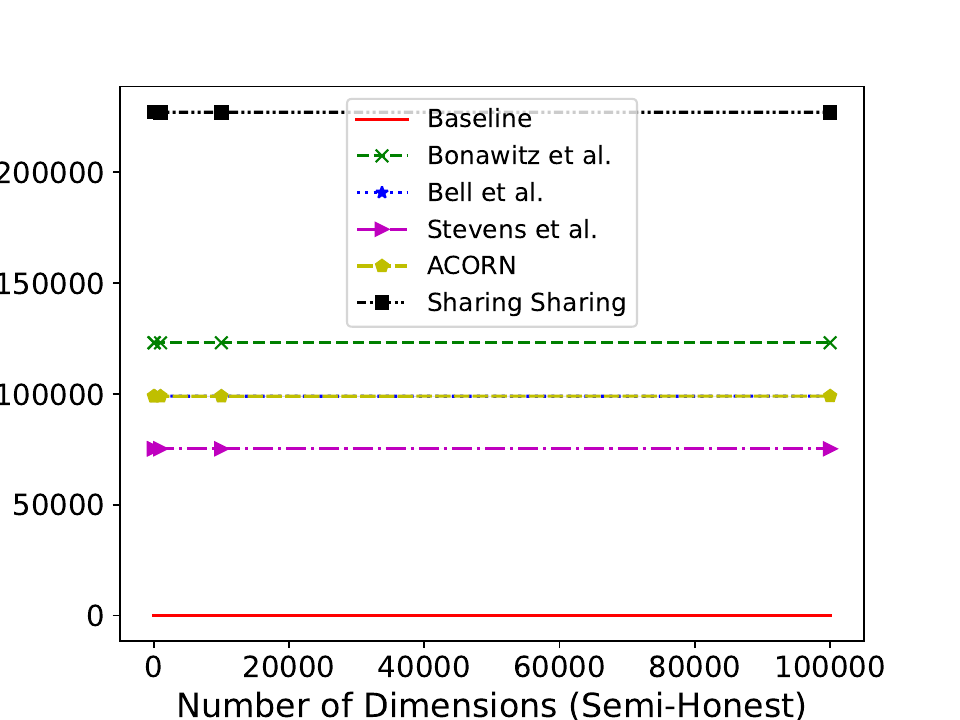}
& \mkgraph{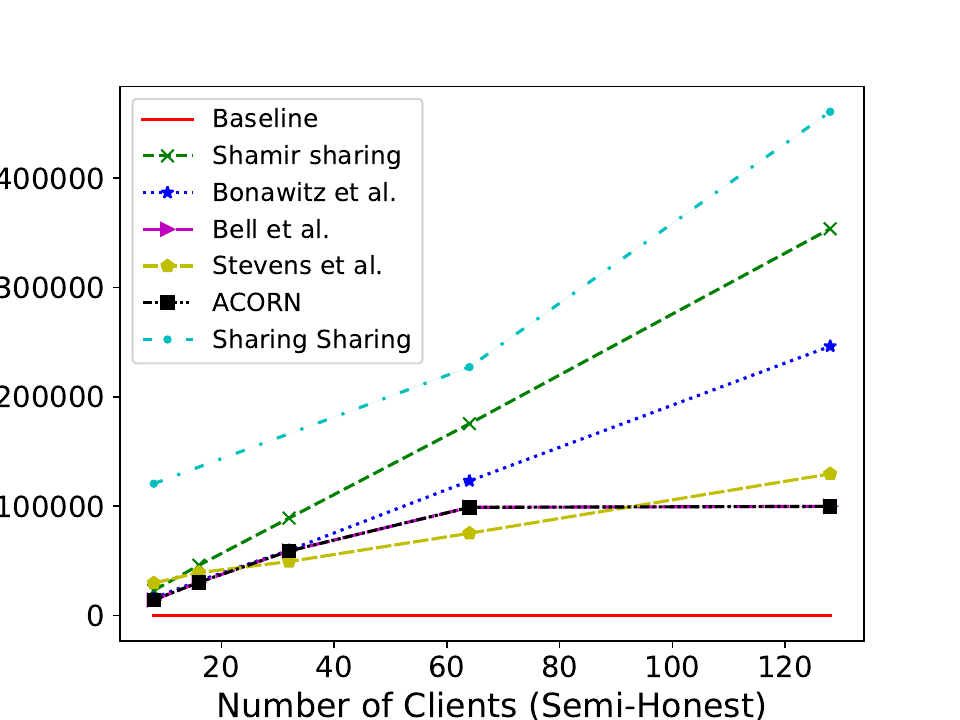}
& \mkgraph{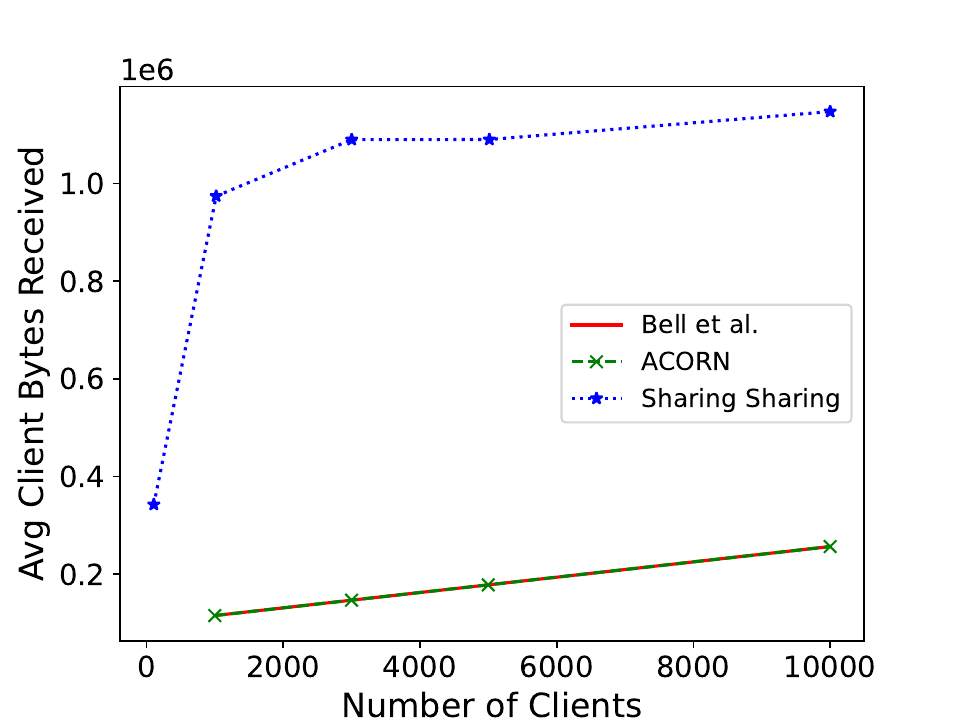}\\
\hline
  \multicolumn{3}{c}{\textbf{Server Bytes Sent (Semi-Honest)}}\\
  \hline
\textbf{\footnotesize (a) Large Vectors (64 clients)} 
& \textbf{\footnotesize (b) Few Clients (length-100 vectors)} 
& \textbf{\footnotesize (c) Many Clients (length-100 vectors)}\\
\mkgraph{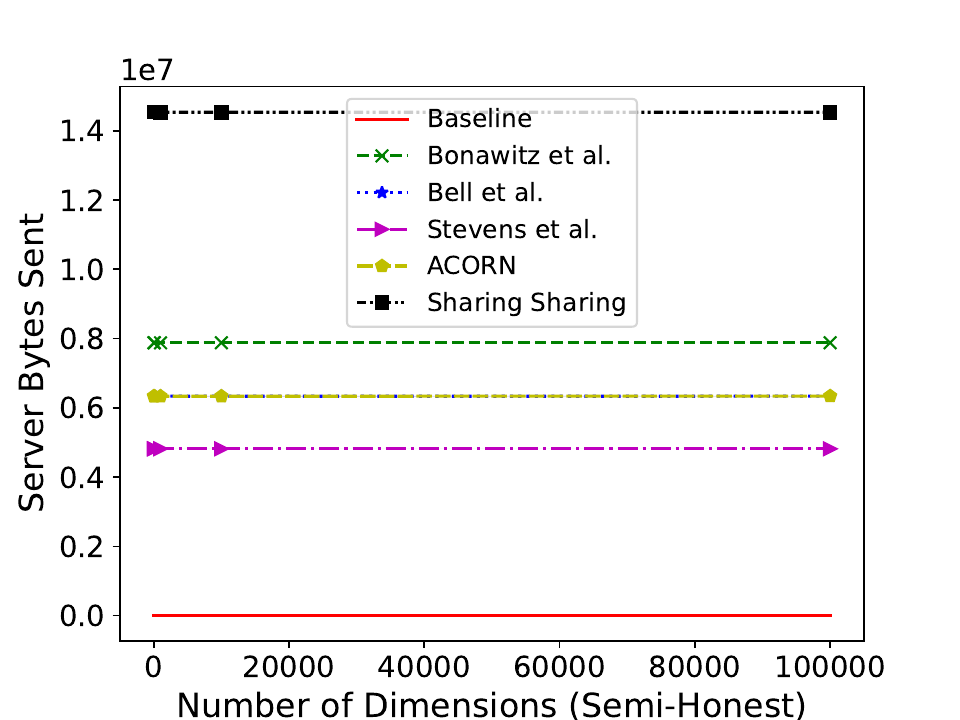}
& \mkgraph{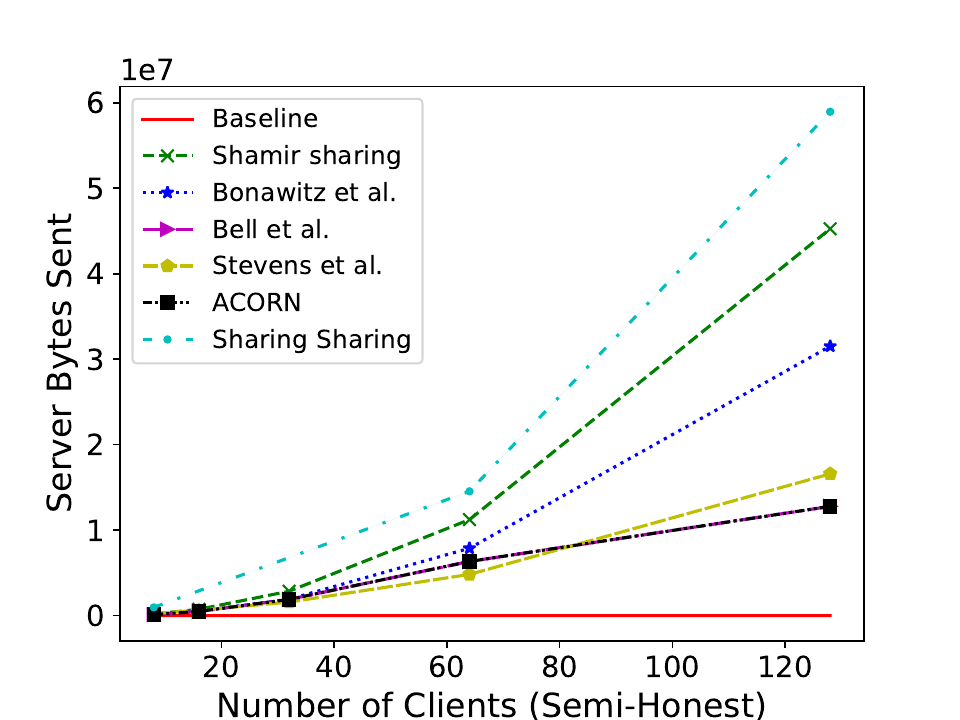}
& \mkgraph{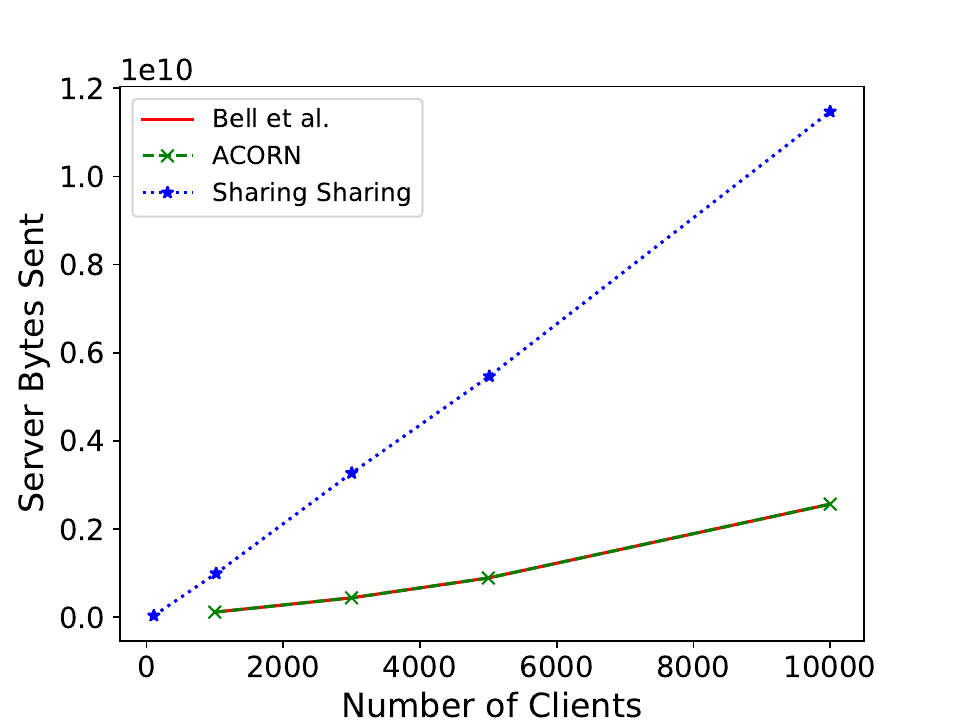}\\
\hline
\end{tabular}
\caption{Additional results for semi-honest protocols.}
\label{fig:results_additional1}
\end{figure*}

Figure~\ref{fig:results_additional1} contains additional results for the semi-honest variants of the case study protocols: the average bytes sent and received by clients, and the bytes sent by the server. These results mirror those presented in Section~\ref{sec:evaluation}---the server sends roughly as much traffic as it receives, and each client sends and receives roughly $1/n$ of the traffic sent and received by the server---because all communication between clients is routed through the server.

\section{Additional Evaluation Results: Malicious Security}
\label{sec:appendix_results2}

\begin{figure*}
\centering
\begin{tabular}{c c}
  \hline
  \multicolumn{2}{c}{\textbf{Avg Client Computation Time (Malicious-Secure)}}\\
  \hline
\textbf{\footnotesize (a) Large Vectors (64 clients)} 
& \textbf{\footnotesize (b) Few Clients (length-100 vectors)} \\
\mkgraph{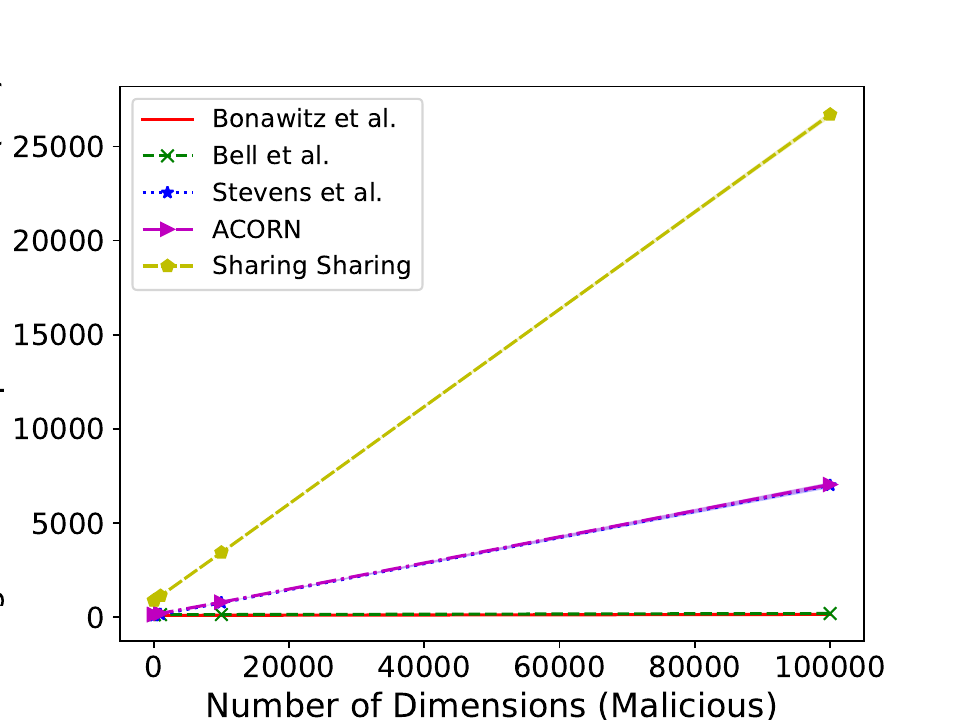}
& \mkgraph{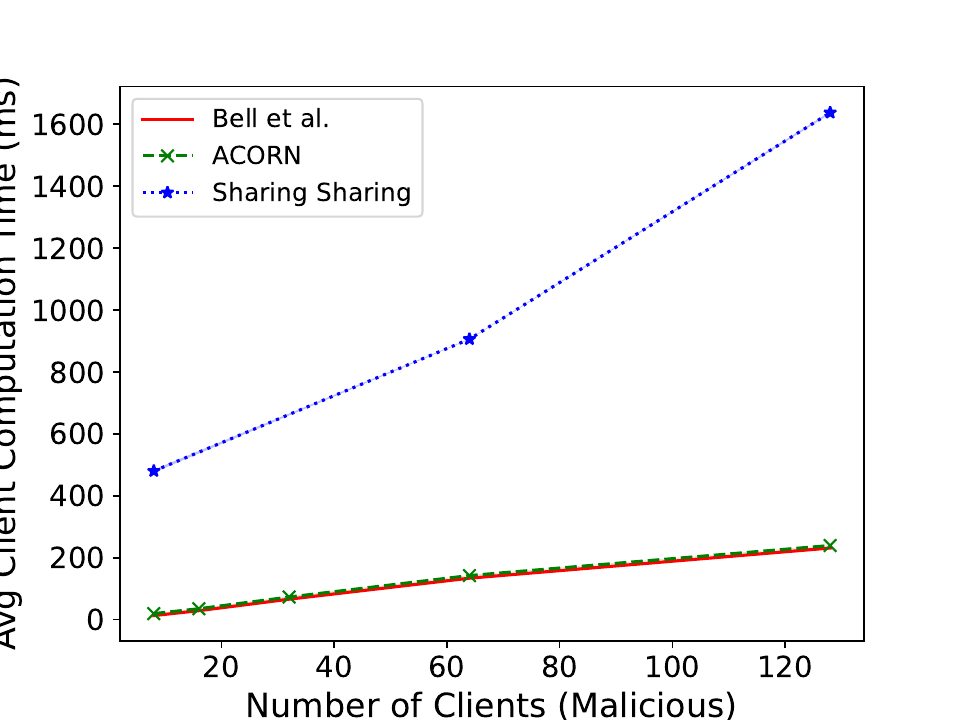}\\
  \hline
  \multicolumn{2}{c}{\textbf{Server Computation Time (Malicious-Secure)}}\\
  \hline
\textbf{\footnotesize (a) Large Vectors (64 clients)} 
& \textbf{\footnotesize (b) Few Clients (length-100 vectors)} \\
\mkgraph{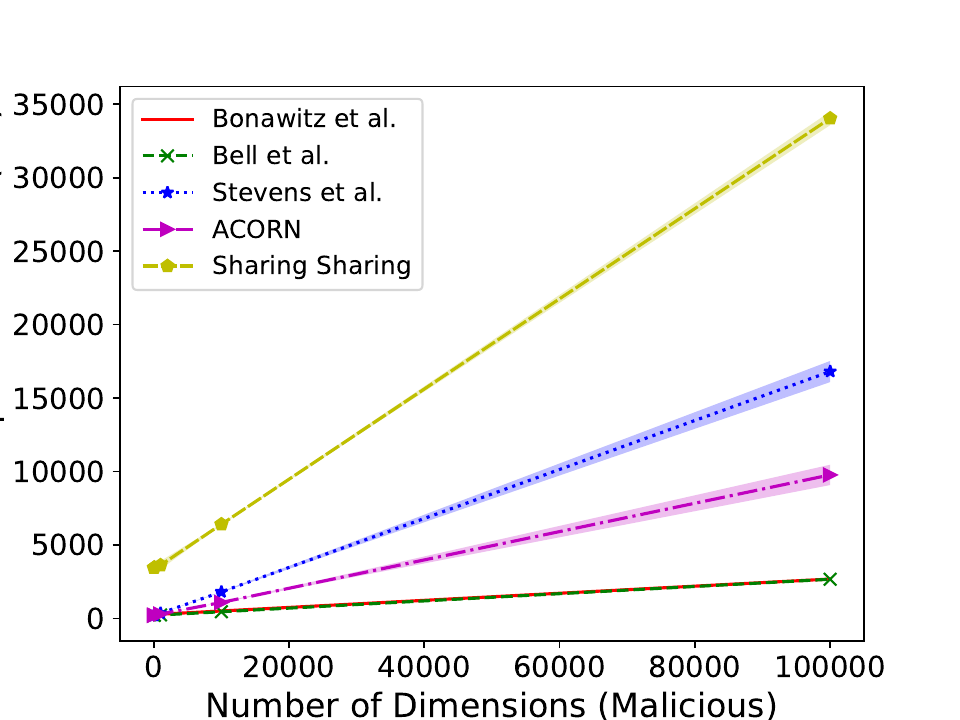}
& \mkgraph{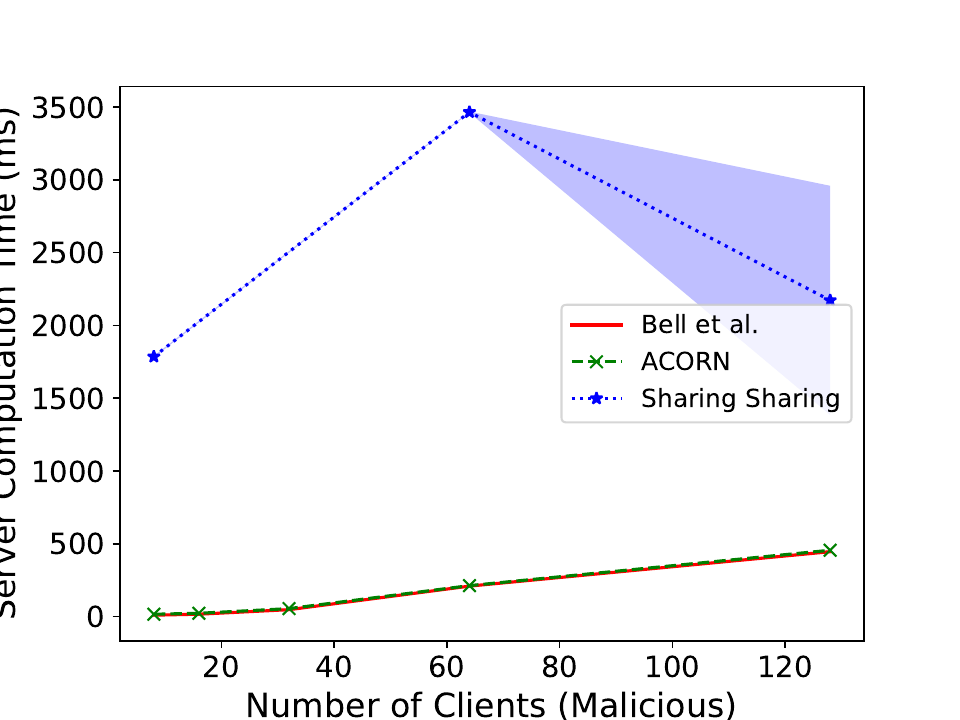}\\
  \hline
  \multicolumn{2}{c}{\textbf{Server Bytes Received (Malicious-Secure)}}\\
  \hline
\textbf{\footnotesize (a) Large Vectors (64 clients)} 
& \textbf{\footnotesize (b) Few Clients (length-100 vectors)} \\
\mkgraph{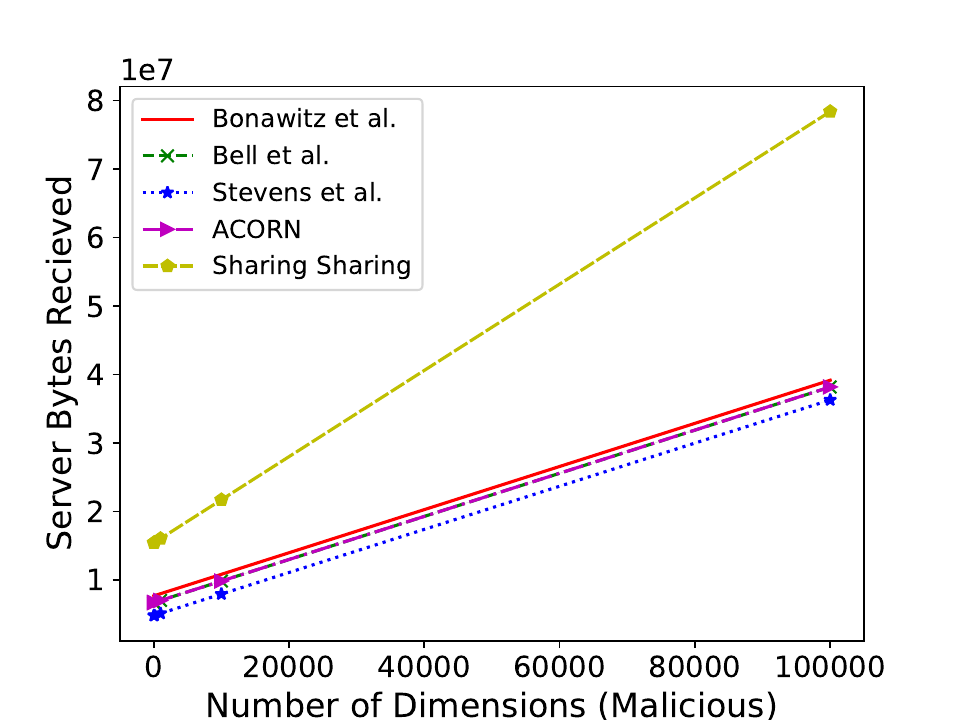}
& \mkgraph{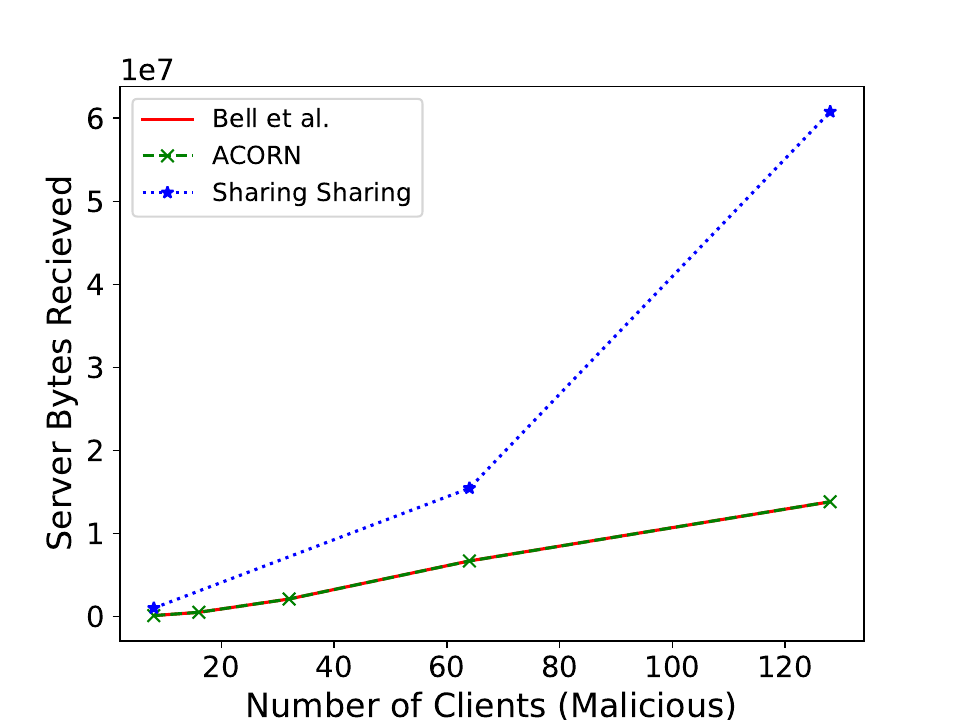}\\
  \hline
  \multicolumn{2}{c}{\textbf{Server Bytes Sent (Malicious-Secure)}}\\
  \hline
\textbf{\footnotesize (a) Large Vectors (64 clients)} 
& \textbf{\footnotesize (b) Few Clients (length-100 vectors)} \\
\mkgraph{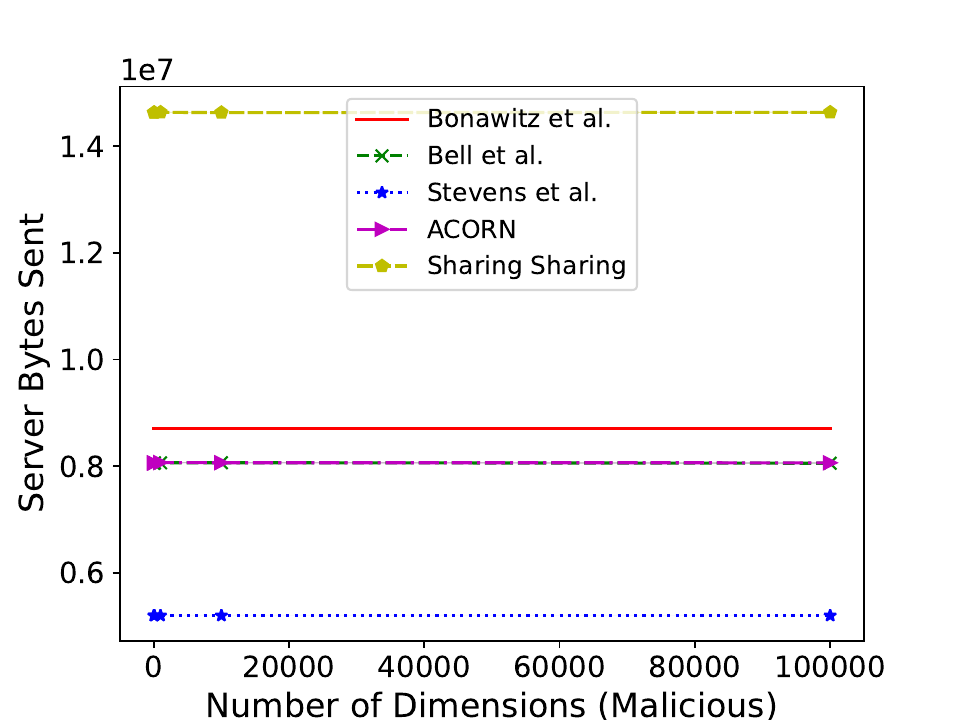}
& \mkgraph{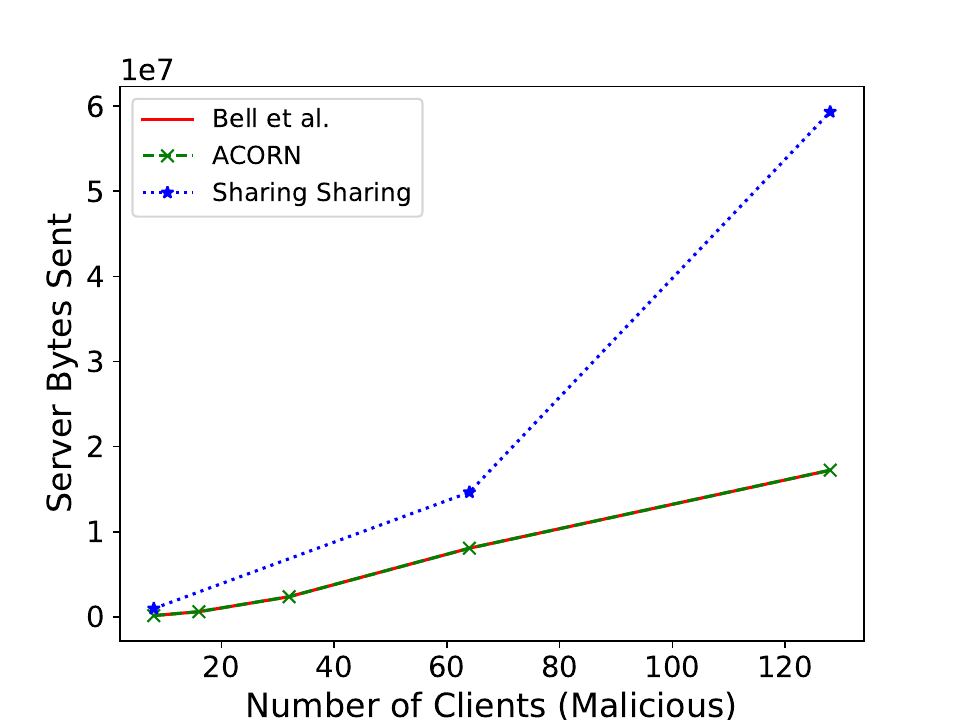}\\
  \hline
\end{tabular}
\caption{Additional results for malicious-secure protocols.}
\label{fig:results_additional2}
\end{figure*}

\begin{figure*}
\centering
\begin{tabular}{c c}
  \hline
  \multicolumn{2}{c}{\textbf{Avg Client Bytes Sent (Malicious-Secure)}}\\
  \hline
\textbf{\footnotesize (a) Large Vectors (64 clients)} 
& \textbf{\footnotesize (b) Few Clients (length-100 vectors)} \\
\mkgraph{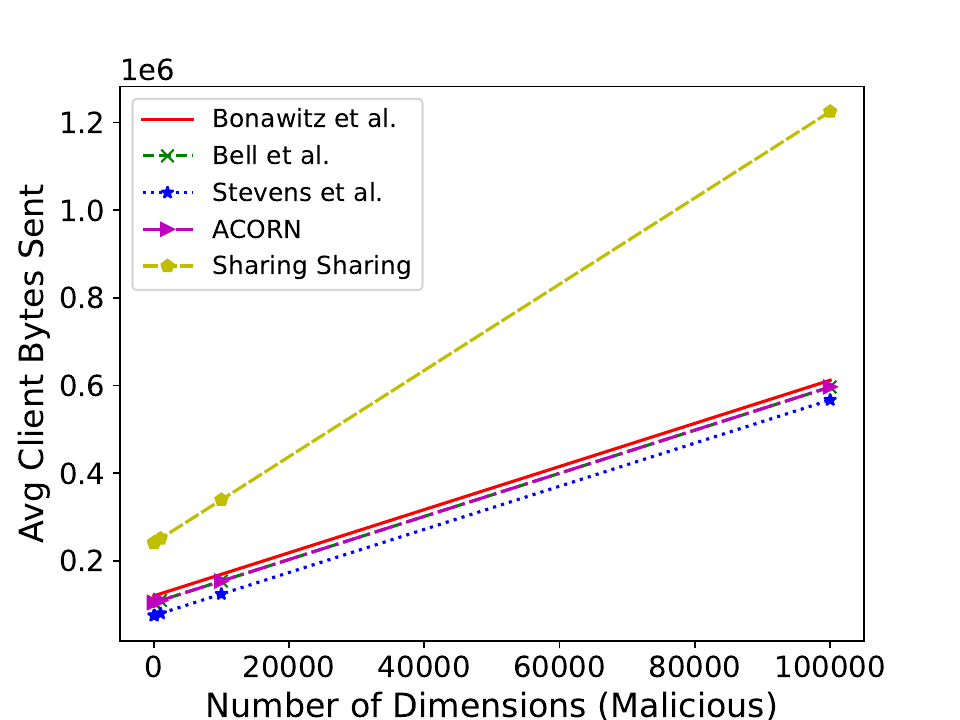}
& \mkgraph{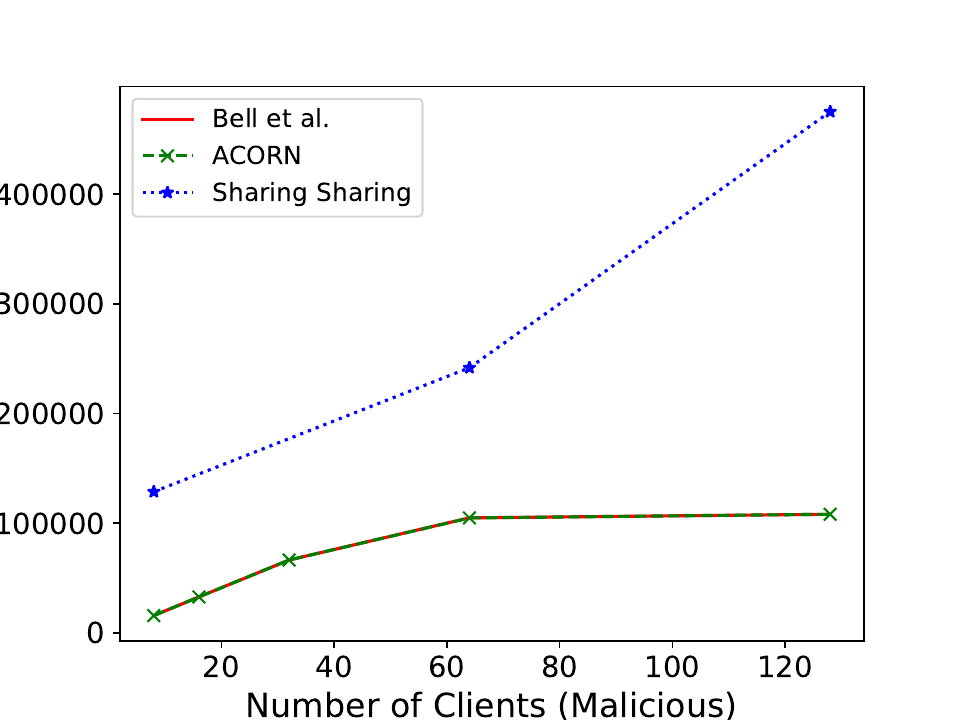}\\
  \hline
  \multicolumn{2}{c}{\textbf{Avg Client Bytes Received (Malicious-Secure)}}\\
  \hline
\textbf{\footnotesize (a) Large Vectors (64 clients)} 
& \textbf{\footnotesize (b) Few Clients (length-100 vectors)} \\
\mkgraph{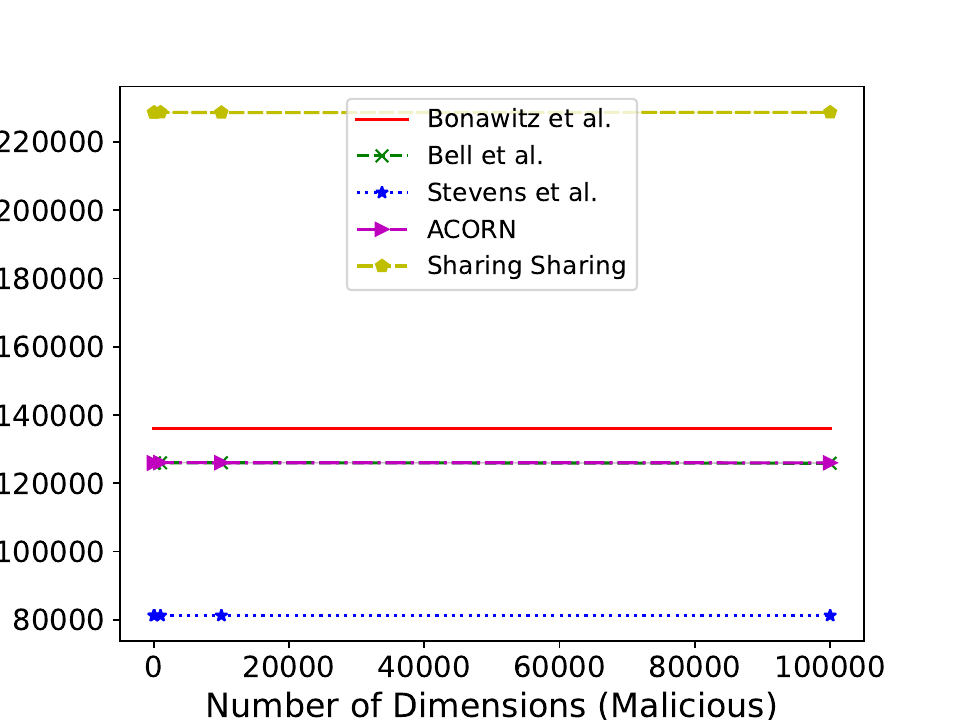}
& \mkgraph{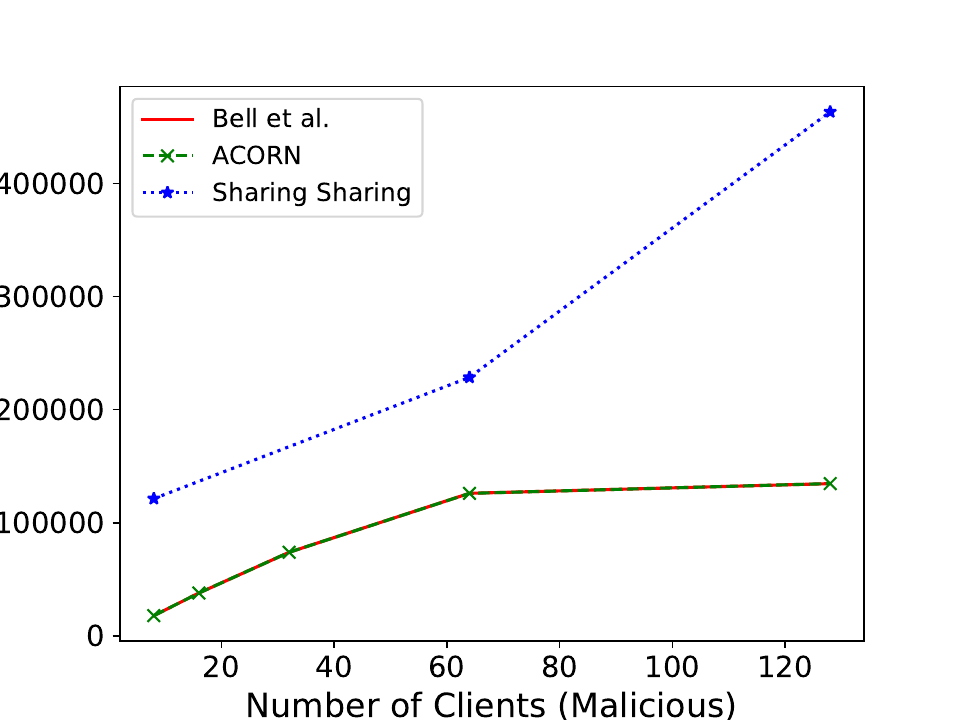}\\
  \hline
\end{tabular}
\caption{Additional results for malicious-secure protocols.}
\label{fig:results_additional3}
\end{figure*}

Figures~\ref{fig:results_additional2} and~\ref{fig:results_additional3} contain additional results for the malicious-secure variants of the case study protocols. These results mirror those for the semi-honest variants presented in Section~\ref{sec:evaluation}, and are included here for completeness.






\end{document}

%% file: imports.tex
\usepackage{amsmath}
\usepackage{amsthm}
\usepackage{amssymb}
\usepackage[frozencache,cachedir=.]{minted}
\usepackage{dcolumn}
\usepackage{booktabs}
\usepackage{tikz}
\usepackage{comment}
\usepackage{graphicx}
\usepackage[ruled, noend, linesnumbered]{algorithm2e}
\usepackage{mdframed}
\let\labelindent\relax
\usepackage{enumitem}
\usepackage[normalem]{ulem}
\usepackage{xcolor}
\usepackage{soul}
\newcommand{\inline}[1]{\mintinline{python}{#1}}
\usepackage{graphicx}
\usepackage{subcaption}
\usepackage{url}
\usepackage{pdfpages}

%% file: macros.tex
\renewcommand{\paragraph}[1]{\noindent\textbf{#1}}
\usetikzlibrary{positioning,shapes,arrows}

\newcommand{\newtext}[1]{#1}

\newcommand{\system}{\textsc{Olympia}\xspace}
\newcommand{\abides}{ABIDES\xspace}

\newcommand{\mkgraph}[1]{\hspace*{-5pt}\includegraphics[width=.33\textwidth, clip, trim={0, 0, 40pt, 35pt}]{#1}\hspace*{-5pt}}